%% file: main.tex
\documentclass[11pt]{article}
\usepackage{amsmath,amssymb,amsfonts,amsthm,bm,mathtools}
\usepackage[margin=1in]{geometry}
\usepackage{setspace,graphicx,booktabs,multirow,longtable,float,caption}
\usepackage[round,authoryear]{natbib}
\usepackage{xcolor,authblk}
\usepackage[T1]{fontenc}
\usepackage{lmodern,microtype}
\usepackage[colorlinks=true,linkcolor=blue,citecolor=blue,urlcolor=blue]{hyperref}
\newtheorem{theorem}{Theorem}
\newtheorem{proposition}{Proposition}
\newtheorem{corollary}{Corollary}

\theoremstyle{definition}\newtheorem{definition}{Definition}
\theoremstyle{remark}\newtheorem{remark}{Remark}

\newcommand{\R}{\mathbb{R}}
\newcommand{\E}{\mathbb{E}}
\newcommand{\I}{\mathbf{1}}
\newcommand{\IG}{\operatorname{IG}}
\newcommand{\Ga}{\operatorname{Gamma}}
\newcommand{\PA}{\operatorname{PALD}}
\newcommand{\bx}{\bm{x}}
\newcommand{\bbeta}{\bm{\beta}}
\newcommand{\Gbar}{\bar{\mathcal{G}}}
\DeclareMathOperator*{\argmin}{arg\,min}

\title{A Bayesian scale-mixture power-asymmetric model for quantile regression with an application to tail-risk forecasting}
\author[1]{Nikita Rawat}
\author[2]{Rishikesh Yadav\thanks{Corresponding author: \texttt{rishikesh@iitmandi.ac.in}}}
\author[3]{Deepak Prajapati}
\affil[1]{Department of Statistics, University of Delhi}
\affil[2]{School of Mathematical and Statistical Sciences, Indian Institute of Technology Mandi}
\affil[3]{Indian Institute of Management Lucknow}
\date{}

\begin{document}
\maketitle

\begin{abstract}
The asymmetric Laplace distribution (ALD) is widely regarded as a working likelihood in Bayesian quantile regression. Its connection to the check loss is useful for quantile estimation, but its fixed shape and exponential tails may limit density forecasts for financial losses. In this paper, we propose a family that preserves the $\tau$-quantile of quantile regression at zero while allowing the shape and tail behaviour of the error distribution to vary. By raising the check loss to a power $r>0$ gives a power asymmetric Laplace distribution (PALD), which includes the ALD and approaches a uniform distribution in the limit as $r \to \infty$. Mixing its scale over an inverse-gamma distribution gives a power-scale mixture of asymmetric Laplace distributions (PSMALD), whose survival exponent depends on the mixing shape while the power affects the body and tail constants. We derive the distribution and quantile functions and moments of the PALD law, the marginal density and tails of PSMALD, and its closed-form limit as $r\to\infty$. We also show that, under misspecification in a homoscedastic location model, the population projection preserves the regression slopes and can shift the intercept, thereby changing the effective quantile level. One-dimensional quadrature allows us to integrate out the latent scales before sampling with the No-U-Turn sampler. We conduct a simulation study to showcase parameter recovery, the weak identification of large powers and the effect of misspecification on coverage. We applied our framework to three financial data and found that in chronological holdouts for Nifty Bank, Ibovespa and the Indian rupee, PSMALD gives the highest mean log predictive density among the three Bayesian fits at high quantile levels. In particular, at the 99\% level, it reduces check loss relative to the ALD in all three datasets, while the comparison at lower levels depends on the model and series. Large fitted scales identify several observations during periods of market stress. 
We have coded our methodology into and \texttt{R} package.
\end{abstract}


%=====================================================================
\section{Introduction}\label{sec:intro}
%=====================================================================
On 23 March 2020, Nifty Bank lost about one sixth of its value in a single session. Eleven days earlier, the Brazilian Ibovespa had fallen by nearly 15\%. These rare but important episodes illustrate why financial risk assessment pays particular attention to the upper tail of the loss distribution. Value-at-risk (VaR) at level $\tau$ is the conditional $\tau$-quantile of loss. Standard VaR backtests assess how often losses exceed the forecast and whether these exceedances cluster \citep{kupiec1995,christoffersen1998}. At the high quantile levels such as $\tau=0.95$ and $0.99$ that are commonly studied, only a small number of episodes can have a substantial effect on estimates. Heavy tails and volatility clustering make these quantiles especially difficult to forecast \citep{cont2001,mcneil2000}. Therefore, the application usually considers both how predictors relate to the conditional quantile and how well the working distribution describes losses around and beyond it. The distinction is important at the 99th percentile, where an adequately calibrated threshold may still be accompanied by a predictive density that assigns too little probability to the largest observed losses. In this work, we examine calibration, check loss and predictive density separately so that each comparison has a clear interpretation.

Quantile regression \citep{koenker1978} addresses the first question by minimising the asymmetric check loss $\sum_i\rho_\tau(y_i-\bx_i^\top\bbeta)$, where $\rho_\tau(u)=u\{\tau-\I(u<0)\}$. It estimates the conditional quantile directly from a criterion that is linear in the residual magnitude on either side of zero. Its inferential theory is well developed \citep{koenker1999,koenker2005}. In finance, this approach underlies the CAViaR models of \citet{engle2004} and other quantile-based methods for forecasting VaR. The same check loss also is a consistent score for evaluating quantile forecasts \citep{gneiting2011}.

The asymmetric Laplace distribution (ALD) provides a likelihood representation of the check loss \citep{yu2001}; its $\tau$-quantile is zero, and maximising its likelihood over the regression coefficients at a fixed scale gives the check-loss estimator. A smart trick comes with representing the ALD as a location-scale mixture of normals, which permits Gibbs sampling with conjugate updates \citep{kozumi2011}. \citet{chernozhukov2003} developed the broader use of criterion-based posteriors for estimation. Under regularity conditions, the ALD posterior concentrates on the true conditional quantile coefficients even when the error law is misspecified \citep{sriram2013}. The uncertainty from this working posterior can require adjustment to obtain nominal frequentist coverage \citep{yang2016}. In a quasi-maximum-likelihood setting, \citet{komunjer2005} characterised the class of tick-exponential densities, including the ALD, that identify conditional quantiles under appropriate conditions.

The usefulness of the ALD for quantile estimation does not directly imply that it gives an adequate predictive density. This is due to the fact that, though its scale may vary, its shape is fixed at a given $\tau$: the density has a cusp at the target quantile and its two tails decay exponentially at rates determined by $\tau$ and the scale. These restrictions matter when the analysis also aims to describe the size of future losses, compare predictive distributions \citep{vehtari2017}, or assess individual observations. Return series of financial data frequently display heavy tails and periods of unusual volatility \citep{cont2001,mcneil2000}. We therefore distinguish two questions throughout the paper: whether a working likelihood estimates the intended quantile and whether its predictive density describes the distribution of losses. Section~\ref{sec:pseudo} makes this distinction explicit under misspecification. In particular, preserving a quantile in the working distribution describes the fitted model; and whether that quantile coincides with the corresponding quantile of the data-generating distribution depends on the population objective induced by the likelihood. This distinction also helps interpret the simulation results when the regression coefficients and the error distribution have different sensitivities to misspecification.

In this work, we have proposed two different strands of work that address these restrictions. First, nonparametric Bayesian approaches use mixtures subject to quantile constraints \citep{kottas2009,reich2010}; they allow flexible error distributions, while parametric families make it easier to study how particular features of the error law affect inference. Parametric approaches include the asymmetric power and asymmetric exponential-power distributions of \citet{komunjer2007} and \citet{zhu2009}, building on the exponential-power family \citep{boxtiao1973}. \citet{bernardi2018} used the skew exponential-power distribution for Bayesian quantile regression, and \citet{yan2017} developed a generalised asymmetric Laplace family in which skewness can vary at a fixed quantile level. Secondly, scale mixtures offer a separate way to accommodate heavy tails \citep{west1987,kotz2001}, including in robust Bayesian quantile regression \citep{wichitaksorn2014}. In this work, we combine an asymmetric power kernel with inverse-gamma mixing, study the quantile level targeted under misspecification, and integrate out the observation-specific scales before sampling. This avoids an update for every observation at each iteration of the sampler. The latent scales can still be summarised after fitting by averaging their conditional moments over the posterior draws, which provides a way to examine individual training observations alongside the estimated coefficients.

Our proposed approach is built upon the quantile-based parameterisation of the asymmetric power distribution \citep{komunjer2007}. The PALD density is proportional to $\exp[-\{\rho_\tau(e)/\sigma\}^r]$. The power $r$ changes the kernel from one with heavier tails than the ALD ($r<1$), through the ALD ($r=1$) and a two-piece Gaussian ($r=2$), to a uniform limit as $r\to\infty$. We then mix the scale over observation-specific  $\lambda_i\sim\IG(a_\lambda,a_\lambda-1)$. In the resulting PSMALD model, $a_\lambda$ fixes the marginal survival exponent, while the power affects the shape near the target quantile and the tail constants. The conditional distribution of each error has probability $\tau$ below zero for every value of its scale, so the mixture preserves $\bx^\top\bbeta_\tau$ as the conditional $\tau$-quantile. Setting $r=1$ and taking the scales to be constant recovers the ALD.

In this work we make five major contributions detailed below.
\begin{enumerate}
\item \emph{A nested family of quantile-preserving working likelihoods.} The PALD and PSMALD distributions, together with a new closed-form limit, PSMALD$_\infty$, form a single family that contains the ALD and in which the kernel power and the marginal survival exponent have separate parameters (Section~\ref{sec:method}).
\item \emph{Distributional properties.} We derive the distribution and corresponding quantile functions, its closed-form moments and the log-concavity region of the PALD law (Theorem~\ref{thm:pa}, Propositions~\ref{prop:pa_moments}--\ref{prop:pa_limit}); a one-dimensional integral representation of the PSMALD density (Theorem~\ref{thm:marginal}); polynomial density tails with exponent $a_\lambda+1$ for every $r$, which separates the roles of the kernel and mixing parameters (Theorem~\ref{thm:tail}); and the limit as $r\to\infty$, a scale mixture of asymmetric uniforms in closed form, which provides a reference point for interpreting large fitted values of $r$ (Theorem~\ref{thm:limit}).
\item \emph{What is estimated when the model is wrong.} We show that every member of the family recovers the slopes of a location-shift quantile regression, while the intercept is shifted by an amount $c^*$ that can be computed from the error law; the forecasts therefore target an effective level $\tau_{\rm eff}$ (Proposition~\ref{prop:pseudo}). This complements the quantile-consistency results of \citet{komunjer2005}.

\item \emph{Marginal computation.} Two exact representations of the PSMALD density, one smooth for small and one for large $r$, allow the latent scales to be integrated out with log-density errors below $10^{-4}$ on the validation grid, so that $(\bbeta,\sigma,r)$ can be sampled with the No-U-Turn (NUTS) sampler \citep{hoffman2014} on several thousand observations; the latent scales are then recovered by Rao--Blackwellisation, averaging conditional moments over posterior draws (Section~\ref{sec:bayes}).
\item \emph{Evidence and software.} We have coded our proposed methodology in a R package \texttt{PowerSMAL} that details the simulation studies of recovery and misspecification (Section~\ref{sec:sim}) and an out-of-sample tail-risk study of three emerging-market series (Section~\ref{sec:case}).
\end{enumerate}

We demonstrate our methodology with an application to the daily losses for the Nifty Bank index, the Brazilian Ibovespa and the Indian rupee. We first test the heavy tailedness of the data by a Generalised Pareto fits at the empirical 95th percentile that give raw tail-shape estimates between 0.17 and 0.29. These descriptive diagnostics motivate the forecast comparison at $\tau=0.90$, $0.95$, $0.97$ and $0.99$. In the fits, PALD often has lower holdout check loss at the lower levels, and PSMALD improves on the ALD at the 99\% level in each series. PSMALD also improves the held-out log predictive density relative to the ALD in all 12 combinations (4 quantile levels and 3 datasets) by up to 0.45 nats per observation. In the training samples, some of the largest fitted scales occur during the 2008 sell-off, the 2013 rupee depreciation and the March 2020 market disruption. The PSMALD fits also give smaller volatility coefficients than the ALD in the reported comparisons. The forecasts use the same information set for each model, comprising the preceding loss, its square and the variability of the preceding 20 returns. The split preserves time order, with coefficients fitted once on the earlier observations and predictors updated through the holdout.

The rest of the paper is organised as follows. Section~\ref{sec:method} recalls the ALD working likelihood and develops the framework and its theory, Section~\ref{sec:bayes} its Bayesian inference. Section~\ref{sec:sim} presents the simulation studies and Section~\ref{sec:case} the case studies. Section~\ref{sec:conclusion} concludes. Proofs, alternative Gibbs samplers and additional results are in the Supplementary Material.

%=====================================================================
\section{The power-asymmetric framework for quantile regression}\label{sec:method}
In this section, we detail our modeling framework, starting with the 
proposed distributions and their connections to quantile regression. 
More explicitly, in Section~\ref{sec:qr_ald}, we briefly recap the 
asymmetric Laplace distribution (ALD), which serves as the starting 
point for our methodological contributions. In Section~\ref{sec:pa}, 
we extend the ALD framework to the Power-ALD (PALD), which closely 
resembles the power asymmetric distributions introduced by 
\cite{komunjer2007}, but is slightly modified to retain the elegant 
properties of the ALD and its advantages for quantile regression. 
In Section~\ref{sec:psmal}, we further extend the PALD framework to 
incorporate observation-specific scale parameters, thereby accounting 
for potential heteroscedasticity in the model. Finally, in Section \ref{sec:pseudo}, we provide the details that connect these distribution with the quantile regression framework and state what each member estimates when the data do not come from it.  

\subsection{Quantile regression and the asymmetric Laplace likelihood}\label{sec:qr_ald}
For a quantile level $\tau\in(0,1)$, the linear quantile regression model is
\begin{equation}\label{eq:qr}
Q_\tau(y_i\mid\bx_i)=\bx_i^\top\bbeta_\tau,\qquad i=1,\dots,n,
\end{equation}
where $\bx_i\in\R^p$ contains an intercept. Writing $y_i=\bx_i^\top\bbeta_\tau+\varepsilon_i$, the model only requires that the $\tau$-quantile of $\varepsilon_i$ given $\bx_i$ is zero. The classical estimator \citep{koenker1978} of $\beta_\tau$ is $\hat\bbeta_\tau=\argmin_{\bbeta}\sum_{i=1}^n\rho_\tau(y_i-\bx_i^\top\bbeta)$ where $\rho_\tau(u)$ is the check-loss function,  defined as
\begin{equation}\label{eq:check}
\rho_\tau(u)=u\{\tau-\I(u<0)\}=\begin{cases}(1-\tau)|u|,&u<0,\\ \tau u,&u\ge0.\end{cases}
\end{equation}
The Bayesian standard \citep{yu2001} uses the asymmetric Laplace likelihood 
\begin{equation}\label{eq:ald}
f_{\rm ALD}(e\mid\tau,\sigma)=\frac{\tau(1-\tau)}{\sigma}\exp\Big\{-\frac{\rho_\tau(e)}{\sigma}\Big\},\qquad\sigma>0,
\end{equation}
whose $\tau$-quantile is zero for every $\sigma$, so that maximising the likelihood over the regression coefficients at fixed scale gives a classical check-loss minimiser. 

Another elegant approach for performing quantile regression is when we assume $V\sim{\rm Exp}(1)$, $Z\sim N(0,1)$, and 
\begin{align}
\label{eq:ald_mixture}
\varepsilon=\sigma\xi_\tau V+\sigma\sqrt{\phi_\tau V}Z, 
\end{align} where $\xi_\tau=(1-2\tau)/\{\tau(1-\tau)\}$ and $\phi_\tau=2/\{\tau(1-\tau)\}$. With this mixture representation $\varepsilon$ has density \eqref{eq:ald}, which forms the basis behind the Gibbs sampler of \citet{kozumi2011}.  Our proposed approach centers around this \eqref{eq:ald_mixture} framework of ALD by keeping its key property e.g., a $\tau$-quantile at zero, but relaxing its two restrictions, a fixed shape with exponential tails and a common scale for all observations, which results in the models of sections \ref{sec:pa} and \ref{sec:psmal}. More explicitly, our proposed framework has two layers. The first replaces the check loss in \eqref{eq:ald} by its power $\rho_\tau^r$, which changes the shape of the body without moving the quantile (Section~\ref{sec:pa}). The second gives every observation its own scale $\sigma\lambda_i$, which produces polynomial tails and permits larger fitted scales for observations with large residuals (Section~\ref{sec:psmal}). The resulting family and its special cases are summarised in Table~\ref{tab:special}; Section~\ref{sec:pseudo} states what each member estimates when the data do not come from it.

\subsection{The power-asymmetric distribution}\label{sec:pa}

\begin{figure}[t]\centering
\includegraphics[width=\textwidth]{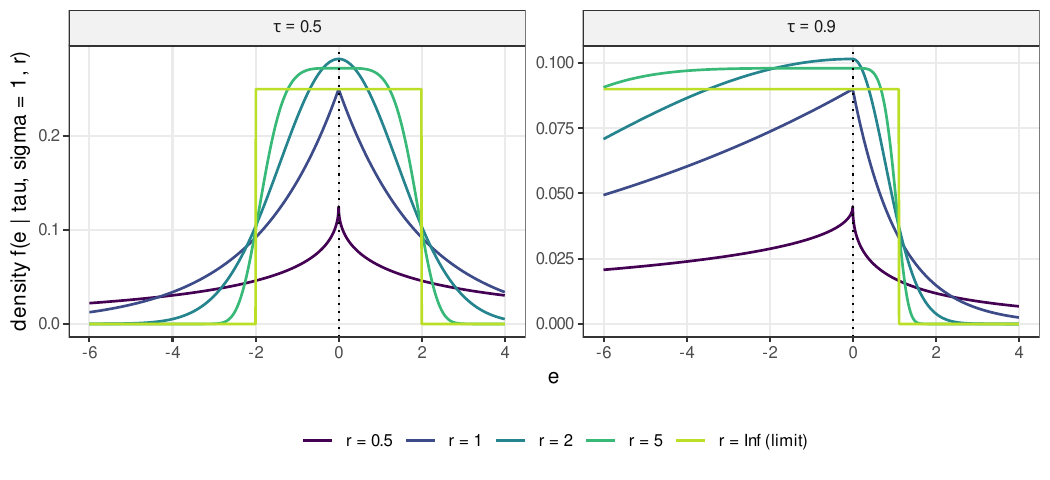}
\caption{Power-asymmetric densities \eqref{eq:pa} with $\sigma=1$ for several powers $r$, at $\tau=0.5$ (left) and $\tau=0.9$ (right). The dotted line marks the $\tau$-quantile, which is zero for every $r$.}\label{fig:pa}
\end{figure}
\begin{definition}[PALD distribution]\label{def:pa}
For $\tau\in(0,1)$, $\sigma>0$ and $r>0$, the power asymmetric Laplace distribution $\PA(\tau,\sigma,r)$ has density
\begin{equation}\label{eq:pa}
f_{\PA}(e\mid\tau,\sigma,r)=\frac{\tau(1-\tau)}{\sigma\,\Gamma(1+1/r)}\exp\Big[-\Big\{\frac{\rho_\tau(e)}{\sigma}\Big\}^r\Big],\qquad e\in\R.
\end{equation}
\end{definition}

Equation~\eqref{eq:pa} is the asymmetric power distribution of \citet{komunjer2007} in a quantile-based scale parameterisation. To see this, set the asymmetry parameter in their standard density to $\alpha=1-\tau$ and the power to $r$. If $\delta_{\alpha,r}$ is their normalising factor inside the exponent, rescaling the standard variable by $\sigma\delta_{\alpha,r}^{1/r}/\{\tau(1-\tau)\}$ gives \eqref{eq:pa}. We use PALD both for the error distribution and for the corresponding regression fit in the numerical comparisons.

Let $\Gbar_\alpha(x)=\Gamma(\alpha,x)/\Gamma(\alpha)$ denote the regularised upper incomplete gamma function and $\Gbar_\alpha^{-1}$ its inverse in $x$.

\begin{theorem}[Distribution and quantile functions]\label{thm:pa}
Let $\varepsilon\sim\PA(\tau,\sigma,r)$. Then \eqref{eq:pa} is a proper density, and
\begin{equation}\label{eq:pa_cdf}
F(e)=\begin{cases}\tau\,\Gbar_{1/r}\big[\{(1-\tau)(-e)/\sigma\}^r\big],&e<0,\\
1-(1-\tau)\,\Gbar_{1/r}\big[\{\tau e/\sigma\}^r\big],&e\ge0,\end{cases}
\qquad
Q(u)=\begin{cases}-\dfrac{\sigma}{1-\tau}\Big\{\Gbar_{1/r}^{-1}\Big(\dfrac{u}{\tau}\Big)\Big\}^{1/r},&u<\tau,\\[2mm]
\dfrac{\sigma}{\tau}\Big\{\Gbar_{1/r}^{-1}\Big(\dfrac{1-u}{1-\tau}\Big)\Big\}^{1/r},&u\ge\tau.\end{cases}
\end{equation}
In particular $F(0)=\tau$ for every $\sigma$ and $r$, so that, if $y_i=\bx_i^\top\bbeta_\tau+\varepsilon_i$ with $\varepsilon_i\sim\PA(\tau,\sigma,r)$, then $Q_\tau(y_i\mid\bx_i)=\bx_i^\top\bbeta_\tau$.
\end{theorem}

The proof (Supplement~\ref{S-app:proofs}) splits the integral at zero and uses $\int_0^\infty e^{-z^r}dz=\Gamma(1+1/r)$. The probability mass is $\tau$ to the left of zero and $1-\tau$ to the right; conditionally on its sign, $\{\rho_\tau(\varepsilon)/\sigma\}^r$ is $\Ga(1/r,1)$-distributed, which gives an exact sampler. Theorem~\ref{thm:pa} is what makes PALD a valid working likelihood for quantile regression: whatever the power $r$, the linear predictor $\bx_i^\top\bbeta_\tau$ keeps its meaning as the conditional $\tau$-quantile, so the shape of the error law can be learnt from the data without changing the target of inference. The closed-form $F$ and $Q$ also give predictive probabilities and quantiles, and hence Value-at-Risk forecasts and probability integral transforms, without numerical integration.

\begin{proposition}[Moments]\label{prop:pa_moments}
For every integer $m\ge1$,
\begin{equation}\label{eq:pa_moments}
\E(\varepsilon^m)=\sigma^m\frac{\Gamma\{(m+1)/r\}}{\Gamma(1/r)}\Big\{\frac{1-\tau}{\tau^m}+(-1)^m\frac{\tau}{(1-\tau)^m}\Big\},
\end{equation}
and $\E|\varepsilon|^m$ is given by the same expression with $(-1)^m$ replaced by $1$. In particular $\E(\varepsilon)=\sigma\,\frac{\Gamma(2/r)}{\Gamma(1/r)}\,\frac{1-2\tau}{\tau(1-\tau)}$, which is zero only for $\tau=1/2$.
\end{proposition}

The proof (Supplement~\ref{S-app:proofs}) integrates each half of \eqref{eq:pa} separately using the gamma integral above. Proposition~\ref{prop:pa_moments} shows that, in contrast with mean regression, the error mean is in general not zero: the regression line tracks the $\tau$-quantile, and the conditional mean of the response lies above or below it by an amount that depends on both $\tau$ and $r$. The conditional mean and variance implied by a fitted quantile regression are therefore available in closed form.

The power $r$ is an interpretable shape parameter (Table~\ref{tab:special} and Figure~\ref{fig:pa}). For $0<r<1$ the tails $\exp(-c|e|^r)$ are heavier than those of the ALD, $r=1$ gives the ALD, $r=2$ gives a two-piece Gaussian kernel, and $r=2$, $\tau=1/2$ gives exactly $N(0,2\sigma^2)$. All moments are finite for every $r$.

\begin{proposition}[Limits and log-concavity]\label{prop:pa_limit}
(a) As $r\to\infty$, $\PA(\tau,\sigma,r)$ converges in total variation to the asymmetric uniform distribution with density $\tau(1-\tau)\sigma^{-1}\I\{-\sigma/(1-\tau)<e<\sigma/\tau\}$, whose $\tau$-quantile is also zero. (b) The density \eqref{eq:pa} is log-concave if and only if $r\ge1$. Hence, for $r\ge1$ and fixed $(\sigma,r)$, the PALD log-likelihood is concave in $\bbeta$, and the conditional posterior of $\bbeta$ under a Gaussian prior is log-concave.
\end{proposition}

The proof is given in Supplement~\ref{S-app:proofs}. Part (a) shows that the PALD family ranges continuously from kernels with heavier tails than the ALD to a bounded kernel, all with the same $\tau$-quantile, so that the data rather than the analyst determine how concentrated the errors are around the quantile line. Part (b) matters for computation: for $r\ge1$ the conditional posterior of $\bbeta$ is unimodal, as it is under the ALD, so in this range the added flexibility does not make the regression coefficients harder to sample.

\begin{table}[t]\centering\small
\caption{Members of the power asymmetric Laplace distribution (PALD) family and of its scale mixture. $\Gamma(1+1/r)$ is the normalising constant of Definition~\ref{def:pa}; $a_\lambda$ is the inverse-gamma shape of the latent scale.}
\label{tab:special}\resizebox{\textwidth}{!}{\begin{tabular}{lll}\toprule
Parameter values & Distribution & Tail of the density\\ \midrule
PALD, $0<r<1$ & heavier-than-Laplace exponential power & $\exp(-c|e|^r)$, stretched exponential\\
PALD, $r=1$ & asymmetric Laplace (ALD) & $\exp(-c|e|)$\\
PALD, $r=2$ & two-piece Gaussian & $\exp(-c e^2)$\\
PALD, $r=2$, $\tau=1/2$ & $N(0,2\sigma^2)$ & $\exp\{-e^2/(4\sigma^2)\}$\\
PALD, $r\to\infty$ & asymmetric uniform on $(-\sigma/(1-\tau),\sigma/\tau)$ & bounded support\\ \midrule
PSMALD, $r=1$ & SMALD (closed form, Corollary~\ref{cor:smal}) & $|e|^{-(a_\lambda+1)}$\\
PSMALD, any $r\in(0,\infty]$ & PSMALD (Theorem~\ref{thm:marginal}) & $|e|^{-(a_\lambda+1)}$ (Theorem~\ref{thm:tail})\\
PSMALD, $r\to\infty$ & PSMALD$_\infty$ (closed form, Theorem~\ref{thm:limit}) & $|e|^{-(a_\lambda+1)}$\\
PSMALD, $\lambda_i\equiv1$ & PALD & as PALD\\
\bottomrule\end{tabular}}\end{table}

\subsection{The PSMALD distribution}\label{sec:psmal}

The PALD family has exponential-type tails for every finite $r$. To allow polynomial tails and observation-specific scale variation, we let each observation have its own latent scale.

\begin{definition}[PSMALD distribution]\label{def:psmal}
$\varepsilon$ has the PSMALD distribution, $\varepsilon\sim{\rm PSMALD}(\tau,\sigma,r,a_\lambda)$, if
\begin{equation}\label{eq:psmal_hier}
\varepsilon\mid\lambda\sim\PA(\tau,\sigma\lambda,r),\qquad\lambda\sim\IG(a_\lambda,b_\lambda),\qquad b_\lambda=a_\lambda-1,\ a_\lambda>1,
\end{equation}
where $\IG(a,b)$ has density $b^a\lambda^{-a-1}e^{-b/\lambda}/\Gamma(a)$, so that $\E(\lambda)=1$.
\end{definition}

The conditional density depends on $(\sigma,\lambda)$ only through $\sigma\lambda$; the restriction $\E(\lambda)=1$ removes this non-identifiability and makes $\sigma$ the global scale and $\lambda$ a relative, observation-specific multiplier. Write $b=b_\lambda$ and $a=a_\lambda$ when no confusion arises.

\begin{theorem}[Quantile preservation and marginal density]\label{thm:marginal}
Let $\varepsilon\sim{\rm PSMALD}(\tau,\sigma,r,a)$ and let $T\sim\Ga(a+1,b)$ (shape, rate). Then $P(\varepsilon\le0)=\tau$, and the marginal density of $\varepsilon$ is
\begin{equation}\label{eq:psmal_marginal}
f_{\rm PSMALD}(e)=\frac{a\,\tau(1-\tau)}{\sigma\,b\,\Gamma(1+1/r)}\;I_r\Big\{\frac{\rho_\tau(e)}{\sigma}\Big\},\qquad
I_r(s)=\E_T\big[\exp\{-(sT)^r\}\big].
\end{equation}
\end{theorem}

The first statement holds because $P(\varepsilon\le0\mid\lambda)=\tau$ for every $\lambda$; the second follows from the substitution $T=1/\lambda$, under which $\lambda^{-1}\IG(\lambda;a,b)\,d\lambda$ becomes a multiple of the $\Ga(a+1,b)$ density. The one-dimensional integral in \eqref{eq:psmal_marginal} is the basis of our computations (Section~\ref{sec:marginal_hmc}). The full proof is in Supplement~\ref{S-app:proofs}. For quantile regression, the first statement means that the latent scales change the spread and the tails of each observation's error without moving its $\tau$-quantile, so that $\bx_i^\top\bbeta_\tau$ remains the conditional quantile under the mixture, while the second makes the likelihood available without the latent scales.

\begin{corollary}[SMALD, $r=1$]\label{cor:smal}
For $r=1$, $I_1(s)=\{b/(b+s)\}^{a+1}$, and
\begin{equation}\label{eq:smal}
f_{\rm SMALD}(e)=\frac{a\,\tau(1-\tau)}{\sigma b}\Big\{1+\frac{\rho_\tau(e)}{\sigma b}\Big\}^{-(a+1)},\qquad
F_{\rm SMALD}(e)=\begin{cases}\tau\{1+(1-\tau)(-e)/(\sigma b)\}^{-a},&e<0,\\ 1-(1-\tau)\{1+\tau e/(\sigma b)\}^{-a},&e\ge0.\end{cases}
\end{equation}
\end{corollary}

The SMALD distribution is a two-piece Pareto (Lomax) law. Its proof (Supplement~\ref{S-app:proofs}) uses the Laplace transform of the gamma distribution. With a closed-form density and distribution function, SMALD gives a heavy-tailed quantile regression likelihood that is as easy to evaluate as the ALD, and it admits the conjugate Gibbs sampler of Supplement~\ref{S-sec:smal_gibbs}. The next result shows that its polynomial tail is not a feature of $r=1$ only.

\begin{theorem}[Tails and moments]\label{thm:tail}
For every $r\in(0,\infty)$, as $|e|\to\infty$,
\begin{equation}\label{eq:tail}
f_{\rm PSMALD}(e)\sim C_r\Big\{\frac{\sigma}{\rho_\tau(e)}\Big\}^{a+1},\qquad C_r=\frac{\tau(1-\tau)\,b^{a}\,\Gamma\{(a+1)/r\}}{\sigma\,\Gamma(a)\,\Gamma(1/r)}.
\end{equation}
For $m\ge0$, $\E|\varepsilon|^m<\infty$ if and only if $m<a$. For integers $1\le m<a$, $\E(\varepsilon^m)=\E(\lambda^m)\,\E_{\PA}(\varepsilon^m)$ with $\E(\lambda^m)=b^m\Gamma(a-m)/\Gamma(a)$ and $\E_{\PA}(\varepsilon^m)$ given by \eqref{eq:pa_moments}.
\end{theorem}

\begin{figure}[!htbp]\centering
\includegraphics[width=\textwidth]{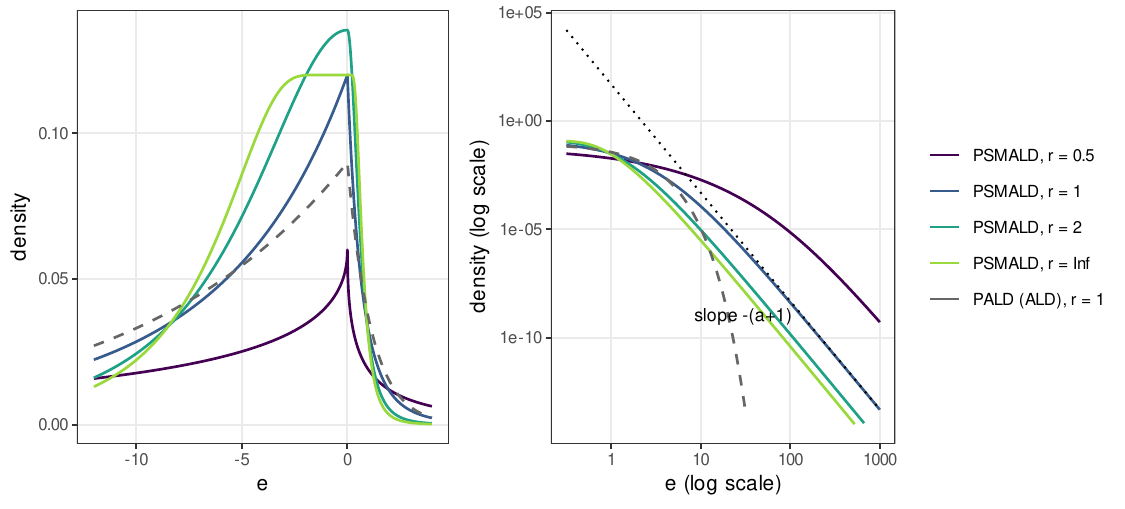}
\caption{PSMALD densities with $\tau=0.9$, $\sigma=1$ and $a_\lambda=4$ for several $r$, and the ALD (dashed). Left: natural scale. Right: right tail on the log--log scale; the dotted line has slope $-(a_\lambda+1)=-5$, the common tail of every PSMALD density (Theorem~\ref{thm:tail}).}\label{fig:psmal}
\end{figure}

Theorem~\ref{thm:tail} separates the two roles of the parameters: the tail index $a$ is set by the mixing law alone, whatever the shape $r$ of the body, while $r$ and $\tau$ only affect the constant. Figure~\ref{fig:psmal} illustrates this: on the log--log scale all PSMALD densities share the slope $-(a+1)$, while the PALD (here ALD) density falls off exponentially. The proof (Supplement~\ref{S-app:proofs}) derives the behaviour of $I_r(s)$ as $s\to\infty$ by monotone convergence and the moments from the inverse-gamma mixing law. For quantile regression of financial losses this is the key property: the probability of a large exceedance of the fitted quantile decays polynomially, at a rate set by $a_\lambda$, rather than exponentially as under the ALD and PALD, while the quantile itself is unaffected (Theorem~\ref{thm:marginal}).

\begin{theorem}[Limit $r\to\infty$]\label{thm:limit}
As $r\to\infty$, ${\rm PSMALD}(\tau,\sigma,r,a)$ converges in total variation to the distribution with density
\begin{equation}\label{eq:psmal_inf}
f_\infty(e)=\frac{a\,\tau(1-\tau)}{\sigma b}\;P\Big\{T<\frac{\sigma}{\rho_\tau(e)}\Big\},\qquad T\sim\Ga(a+1,b),
\end{equation}
which we call PSMALD$_\infty$. It is the law of $\varepsilon$ when $\varepsilon\mid\lambda$ is uniform on $(-\sigma\lambda/(1-\tau),\sigma\lambda/\tau)$ and $\lambda\sim\IG(a,b)$; its $\tau$-quantile is zero, its density has a flat peak at zero (all derivatives of positive order of $f_\infty$ vanish at $e=0$), and its tail is \eqref{eq:tail} with $\Gamma\{(a+1)/r\}/\Gamma(1/r)$ replaced by its limit $1/(a+1)$.
\end{theorem}

The proof (Supplement~\ref{S-app:proofs}) uses dominated convergence and Scheff\'e's lemma. The limit completes the family at $r=\infty$ with a closed-form density, so that the whole range of kernel powers, from sharply peaked to flat-topped, is covered while the $\tau$-quantile and the polynomial tail are preserved.

\begin{remark}\label{rem:khintchine}
The limit has a uniform conditional kernel. Finite powers can give marginal densities close to this limit, which makes large values of $r$ difficult to distinguish from one another. The mixing distribution still restricts the shape of the marginal density.
\end{remark}

The models are nested: PSMALD reduces to PALD when $\lambda_i\equiv1$, to SMALD when $r=1$, to ALD when both hold, and to PSMALD$_\infty$ in the limit $r\to\infty$.

\subsection{The quantile regression model and its target under misspecification}\label{sec:pseudo}

The regression model is $y_i=\bx_i^\top\bbeta_\tau+\varepsilon_i$ with independent errors from one of the working laws ALD, PALD, PSMALD or PSMALD$_\infty$ (or SMALD). By Theorems~\ref{thm:pa}--\ref{thm:limit}, $\bx_i^\top\bbeta_\tau$ is then the conditional $\tau$-quantile of $y_i$ under each of them. In practice the working law is never exactly correct, and the question is what the posterior estimates. Under standard regularity conditions the posterior concentrates on the minimiser of the Kullback--Leibler divergence from the data-generating process to the working model \citep{white1982,kleijn2012}. The next proposition identifies this pseudo-true value.

\begin{proposition}[Pseudo-true parameters]\label{prop:pseudo}
Let $(y_i,\bx_i)$ be i.i.d., $y_i=\bx_i^\top\bbeta_0+\varepsilon_i$ with $\varepsilon_i$ independent of $\bx_i$ and $F_\varepsilon$ continuous and strictly increasing on its support, with $F_\varepsilon(0)=\tau$, where $\bx_i$ contains an intercept and $\E(\bx_i\bx_i^\top)$ is finite and positive definite. Let $\{f_\theta\}$ be a working family of error densities. Assume the expected log likelihoods in the argument are well defined and that $h(c,\theta)=\E\log f_\theta(\varepsilon-c)$ has a unique maximiser $(c^*,\theta^*)$. Then the pseudo-true value of $(\bbeta,\theta)$ is $(\bbeta_0+c^*\bm{e}_1,\theta^*)$, where $\bm{e}_1$ is the first coordinate vector. Thus (a) the slopes are always recovered; (b) the fitted quantile $\bx^\top(\bbeta_0+c^*\bm{e}_1)$ is the conditional quantile of level $\tau_{\rm eff}=F_\varepsilon(c^*)$; (c) for the ALD, $c^*=0$ and $\tau_{\rm eff}=\tau$ when the expected check loss is finite; (d) for the other working laws, $c^*=0$ when the error law belongs to the working family, but in general $c^*\ne0$.
\end{proposition}
\begin{proof}[Proof sketch]
For any $\bbeta$, with $\bm\delta=\bbeta-\bbeta_0$, independence gives $\E\log f_\theta(y-\bx^\top\bbeta)=\E_{\bx}\{h(\bx^\top\bm\delta,\theta)\}\le h(c^*,\theta^*)$, with equality if and only if $\bx^\top\bm\delta=c^*$ almost surely, that is, $\bm\delta=c^*\bm{e}_1$. For the ALD, $h(c,\sigma)=\log\{\tau(1-\tau)/\sigma\}-\E\rho_\tau(\varepsilon-c)/\sigma$, maximised in $c$ at the $\tau$-quantile of $\varepsilon$, which is zero. Details are in Supplement~\ref{S-app:proofs}.
\end{proof}

For quantile regression, Proposition~\ref{prop:pseudo} describes the price of flexibility: in a location model the pseudo-true slopes of a misspecified PALD or PSMALD fit still equal the true slopes, but its intercept, and hence its fitted quantile, may correspond to a level $\tau_{\rm eff}$ different from $\tau$, whereas the ALD is protected against this shift.

Proposition~\ref{prop:pseudo} concerns the population projection for a location model. The connection to Bayesian inference requires posterior concentration at that projection. With covariate-dependent error scales, the independence argument in the proof no longer applies and the flexible likelihoods may also shift the slopes. The ALD population objective continues to identify a correctly specified linear conditional quantile under the usual integrability and identification conditions \citep{komunjer2005,sriram2013}.

\begin{figure}[!htbp]\centering
\includegraphics[width=\textwidth]{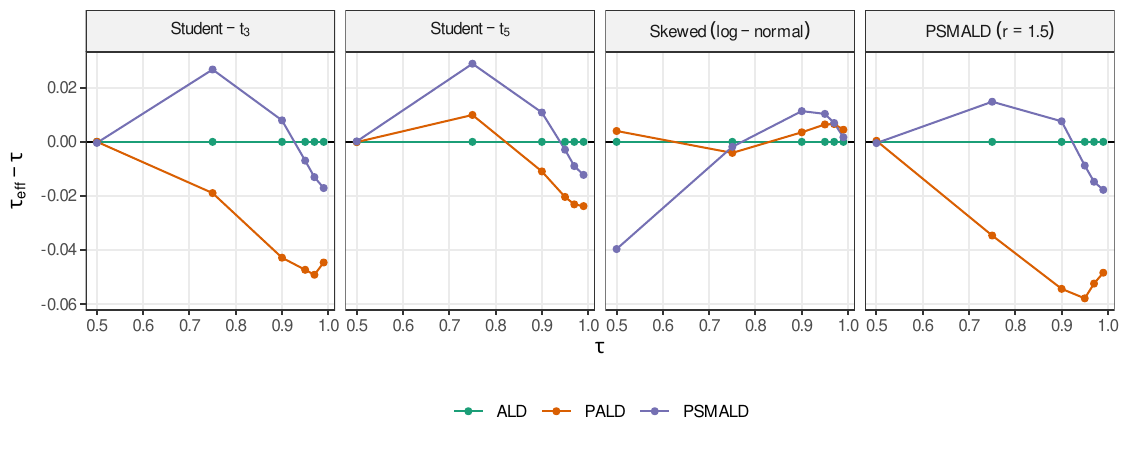}
\caption{Effective quantile level $\tau_{\rm eff}-\tau$ targeted by each working likelihood when the errors follow the stated law (Proposition~\ref{prop:pseudo}); computed from $2\times10^5$ draws. The ALD line is zero for every law.}\label{fig:pseudo}
\end{figure}

Figure~\ref{fig:pseudo} illustrates the intercept shift for four error laws. For Student-$t_3$ errors at $\tau=0.99$, the numerical effective levels are 0.945 for PALD and 0.973 for PSMALD. At several upper levels, the unconstrained PSMALD optimisation returns powers much greater than 50 (Supplementary Table~\ref{S-tab:pseudo}). Their effective levels illustrate the large-power regime. The optimisation alone cannot establish where the population objective attains its supremum. The Bayesian fits described below restrict $r$ to $[0.2,50]$.

%=====================================================================
\section{Bayesian inference}\label{sec:bayes}
%=====================================================================
\subsection{Prior distributions and posterior}\label{sec:prior}
We standardise the response and the non-intercept covariates on the training data and report all results on the original scale. On the standardised scale we use independent priors
\begin{equation}\label{eq:priors}
\bbeta_\tau\sim N_p(\bm0,5^2I_p),\qquad\log\sigma\sim N(-1,1.5^2),\qquad r\sim\Ga(2,0.5)\ \text{truncated to }[0.2,50],
\end{equation}
and fix $a_\lambda=4$, $b_\lambda=3$, so that $\E(\lambda_i)=1$, the tail index of PSMALD is 4 and its variance is finite. The untruncated gamma prior on $r$ has mean 4 and is weakly informative over the whole range of shapes in Table~\ref{tab:special}. With $\bm\lambda=(\lambda_1,\dots,\lambda_n)$ and $e_i=y_i-\bx_i^\top\bbeta_\tau$, the joint posterior of the PSMALD model is
\begin{equation}\label{eq:posterior}
p(\bbeta_\tau,\sigma,r,\bm\lambda\mid\bm y)\propto p(\bbeta_\tau)p(\sigma)p(r)\prod_{i=1}^n\frac{\tau(1-\tau)}{\sigma\lambda_i\Gamma(1+1/r)}\exp\Big[-\Big\{\frac{\rho_\tau(e_i)}{\sigma\lambda_i}\Big\}^r\Big]\IG(\lambda_i;a_\lambda,b_\lambda).
\end{equation}
The ALD, PALD and PSMALD$_\infty$ posteriors are obtained from the corresponding densities in Section~\ref{sec:method}.

\subsection{Full conditional distributions}\label{sec:fullcond}
Several full conditionals of \eqref{eq:posterior} are available in closed form after the reparameterisation $\eta=\sigma^r$. With an alternative prior in which $\eta\sim\IG(a_0,b_0)$ independently of $r$,
\begin{equation}\label{eq:eta_full}
\eta\mid\bbeta_\tau,r,\bm\lambda,\bm y\sim\IG\Big(a_0+\frac nr,\ b_0+\sum_{i=1}^n\Big\{\frac{\rho_\tau(e_i)}{\lambda_i}\Big\}^r\Big),
\end{equation}
while $p(\lambda_i\mid\cdot)\propto\lambda_i^{-(a_\lambda+2)}\exp\{-b_\lambda/\lambda_i-(\rho_\tau(e_i)/(\sigma\lambda_i))^r\}$, $p(\bbeta_\tau\mid\cdot)\propto p(\bbeta_\tau)\exp\{-\sum_i(\rho_\tau(e_i)/\lambda_i)^r/\eta\}$ and $p(r\mid\cdot)\propto p(r)\Gamma(1+1/r)^{-n}\eta^{-n/r}\exp\{-\sum_i(\rho_\tau(e_i)/\lambda_i)^r/\eta\}$ are non-standard. A Metropolis-within-Gibbs sampler based on these conditionals, and a fully conjugate Gibbs sampler for the SMALD special case that extends the mixture representation of \citet{kozumi2011}, are given in Supplement~\ref{S-app:gibbs}. These schemes update the $n$ latent scales separately. The reported fits instead use the marginal sampler with the priors in \eqref{eq:priors}.

\subsection{Integrating out the latent scales}\label{sec:marginal_hmc}
By Theorem~\ref{thm:marginal}, the latent scales can be integrated out analytically up to the one-dimensional integral $I_r(s)$, and the marginal posterior of $(\bbeta_\tau,\sigma,r)$ is sampled with the No-U-Turn sampler \citep{hoffman2014} in Stan \citep{carpenter2017}. For SMALD and PSMALD$_\infty$ the marginal densities \eqref{eq:smal} and \eqref{eq:psmal_inf} are in closed form. For PSMALD we evaluate $I_r(s)$ with one of two exact representations.

\begin{proposition}[Two representations of $I_r$]\label{prop:quad}
Let $A=\log T$, $T\sim\Ga(a+1,b)$, with density $p_A$ and distribution function $F_A$, and let $s>0$. Then $I_r(s)=\E[\exp\{-(se^{A})^r\}]$ and, for every $y_0\in\R$,
\begin{equation}\label{eq:quad2}
I_r(s)=F_A\Big(\frac{y_0}{r}-\log s\Big)+\int_{y_0}^{\infty}e^{-e^{y}}\,p_A\Big(\frac yr-\log s\Big)\frac{dy}{r}-\int_{-\infty}^{y_0}\big(1-e^{-e^{y}}\big)p_A\Big(\frac yr-\log s\Big)\frac{dy}{r}.
\end{equation}
The last term lies between 0 and $e^{y_0}F_A(y_0/r-\log s)$, so it is negligible relative to the first for $y_0=-18$.
\end{proposition}

The first representation is a smooth expectation when $r$ is small and is evaluated by the trapezoid rule on a fixed grid of 70 points in $A$. For large $r$ the integrand $\exp\{-(se^A)^r\}$ becomes a step function in $A$, which the fixed grid may resolve poorly; the substitution $y=r(\log s+A)$ fixes the position of the step, and \eqref{eq:quad2} with $y_0=-18$ is evaluated by Simpson's rule on 61 points over $[y_0,\min\{3.8,r(3+\log s)\}]$. We switch at $r=2.5$. On the validation grid with $s\in[10^{-5},10^3]$ and selected $r\in[0.2,50]$, the maximum absolute error of $\log f_{\rm PSMALD}$ is below $10^{-4}$ (Supplementary Figure~\ref{S-fig:quad}). A naive fixed grid in $\log\lambda$, by contrast, has errors of order $10^{-1}$ when $r\ge10$, which is enough to distort the posterior of $r$.

The approximation is piecewise smooth and may differ slightly across the change of quadrature rule. The posterior is sampled with four chains of 750 warm-up and 750 sampling iterations in the case studies, a diagonal mass matrix and target acceptance rate 0.95, using initial values based on the check-loss solution. Convergence diagnostics are reported in the Supplementary Material (Section~\ref{S-app:computation} and Table~\ref{S-tab:convergence}).

\subsection{Posterior of the latent scales}\label{sec:lambda}
Given $(\bbeta_\tau,\sigma,r)$, the $\lambda_i$ are independent with $p(\lambda_i\mid\cdot)\propto\lambda_i^{-(a+2)}e^{-b/\lambda_i}\exp\{-(s_i/\lambda_i)^r\}$, $s_i=\rho_\tau(e_i)/\sigma$. Writing $I_r^{(m)}(s)=\E\exp\{-(sT_m)^r\}$ with $T_m\sim\Ga(m,b)$, the same substitution as in Theorem~\ref{thm:marginal} gives
\begin{equation}\label{eq:lambda_post}
\E(\lambda_i\mid\bbeta_\tau,\sigma,r,\bm y)=\frac ba\,\frac{I^{(a)}_r(s_i)}{I^{(a+1)}_r(s_i)},\qquad
\E(\lambda_i^2\mid\cdot)=\frac{b^2}{a(a-1)}\frac{I^{(a-1)}_r(s_i)}{I^{(a+1)}_r(s_i)}.
\end{equation}
Averaging \eqref{eq:lambda_post} over posterior draws gives a Rao--Blackwellised estimate of $\E(\lambda_i\mid\bm y)$. Observations with $\E(\lambda_i\mid\bm y)\gg1$ are those the model treats as outliers. In the financial application these are crisis days.

\subsection{Model assessment and forecast evaluation}\label{sec:eval}
The density fit of the working likelihoods is compared by the expected log predictive density estimated by Pareto-smoothed leave-one-out cross-validation (LOOIC) and WAIC \citep{vehtari2017,watanabe2010} as pointwise diagnostics on the training data, and by the mean log predictive density on the test data, a strictly proper scoring rule \citep{gneiting2007}. Quantile forecasts $\hat q_t=\bx_t^\top\hat\bbeta_\tau$, with $\hat\bbeta_\tau$ the posterior mean, are evaluated by the check loss, which is consistent for the $\tau$-quantile \citep{gneiting2011}, and by the exceedance indicators $h_t=\I(y_t>\hat q_t)$, whose mean should be $1-\tau$. We report the exceedance ratio $\bar h/(1-\tau)$ and the unconditional-coverage test of \citet{kupiec1995}, the independence test of \citet{christoffersen1998} and the dynamic quantile (DQ) test of \citet{engle2004}, which regresses $h_t-(1-\tau)$ on a constant, four lagged exceedances and $\hat q_t$. Check losses of two models are compared by the test of \citet{diebold1995} with the Newey--West variance \citep{newey1987}. The tests use nominal significance levels. In the time-series application, the chronological holdout supplies the direct forecast comparison because the training design includes lagged outcomes.

%=====================================================================
\section{Simulation studies}\label{sec:sim}
%=====================================================================
The simulations address two questions. Does the Bayesian procedure of Section~\ref{sec:bayes} recover the parameters of the proposed model, including the power $r$, when the model is correct (Section~\ref{sec:sim_recovery})? And how do the members of the framework compare with one another and with the ALD when the error law is or is not in their family, and does Proposition~\ref{prop:pseudo} predict what they estimate (Section~\ref{sec:sim_misspec})? Both studies use $y_i=1+2x_i+s(x_i)\varepsilon_i$ with $x_i\sim U(0,2)$ and errors whose $\tau$-quantile is zero, so that the true conditional quantile is $Q(\tau\mid x)=1+2x$ in every design. All fits use the priors \eqref{eq:priors} and two chains, with 500 warm-up and 500 sampling iterations in the first study and 750 of each in the second.

\subsection{Recovery of the truth under the PSMALD model}\label{sec:sim_recovery}
Data are generated from the PSMALD model, $\varepsilon_i\sim{\rm PSMALD}(\tau,1,r,4)$ and $s(x)=1$, for three powers, $r\in\{0.75,1.5,4\}$ (a heavier-than-Laplace body, a moderately rounded body and a nearly flat-topped one), three quantile levels, $\tau\in\{0.25,0.5,0.9\}$, and two sample sizes, $n\in\{300,1000\}$, with $R=40$ replications per combination, 720 fits in all. Table~\ref{tab:sim_recovery} reports the bias and root mean squared error (RMSE) of the posterior mean and the coverage (CP) of 95\% credible intervals for the slope $\beta_1$ and the power $r$; Supplementary Table~\ref{S-tab:sim_recovery_full} gives the same for $\beta_0$ and $\sigma$. 

\begin{table}[t]\centering\small
\caption{Simulation study 1: recovery of the slope $\beta_1=2$ and the power $r$ by the Bayesian PSMALD fit when the model is correct ($\sigma=1$, $a_\lambda=4$, $R=40$ replications per cell). Bias and RMSE of the posterior mean; CP: coverage of 95\% credible intervals; E/P: ratio of the empirical SD of the posterior mean across replications to the average posterior SD; $^\dagger$: CP below the Monte Carlo limit 0.882. Supplementary Table~\ref{S-tab:sim_recovery_full} gives the same for $\beta_0$ and $\sigma$.}
\label{tab:sim_recovery}\setlength{\tabcolsep}{4.5pt}
\begin{tabular}{rrr rrrr rrrr}\toprule
 & & & \multicolumn{4}{c}{$\beta_1$} & \multicolumn{4}{c}{$r$}\\ \cmidrule(lr){4-7}\cmidrule(lr){8-11}
$n$ & $r$ & $\tau$ & Bias & RMSE & CP & E/P & Bias & RMSE & CP & E/P\\ \midrule
300 & 0.75 & 0.25 & 0.020 & 0.230 & 0.90 & 1.01 & 0.068 & 0.159 & 0.93 & 1.14\\
 & 0.75 & 0.50 & 0.031 & 0.171 & 0.95 & 0.80 & 0.075 & 0.134 & 0.97 & 0.89\\
 & 0.75 & 0.90 & 0.018 & 0.403 & 0.95 & 1.08 & 0.100 & 0.167 & 0.88$^\dagger$ & 1.03\\
 & 1.5 & 0.25 & -0.011 & 0.152 & 1.00 & 0.95 & 0.601 & 0.837 & 0.90 & 0.67\\
 & 1.5 & 0.50 & 0.026 & 0.146 & 0.95 & 1.07 & 0.568 & 0.982 & 0.93 & 0.97\\
 & 1.5 & 0.90 & 0.034 & 0.214 & 1.00 & 0.91 & 0.563 & 0.881 & 0.93 & 0.85\\
 & 4 & 0.25 & 0.015 & 0.131 & 0.97 & 1.01 & 0.870 & 1.531 & 1.00 & 0.55\\
 & 4 & 0.50 & -0.008 & 0.109 & 0.93 & 0.97 & 0.671 & 1.275 & 1.00 & 0.49\\
 & 4 & 0.90 & -0.038 & 0.173 & 0.97 & 0.87 & 0.587 & 1.437 & 1.00 & 0.61\\
\midrule
1,000 & 0.75 & 0.25 & 0.003 & 0.088 & 0.97 & 0.75 & 0.023 & 0.071 & 0.90 & 1.12\\
 & 0.75 & 0.50 & -0.010 & 0.101 & 0.95 & 1.04 & 0.016 & 0.062 & 1.00 & 1.01\\
 & 0.75 & 0.90 & 0.012 & 0.170 & 0.95 & 0.98 & 0.007 & 0.061 & 0.95 & 1.06\\
 & 1.5 & 0.25 & -0.030 & 0.092 & 0.95 & 1.05 & 0.079 & 0.178 & 1.00 & 0.77\\
 & 1.5 & 0.50 & 0.003 & 0.076 & 0.93 & 1.05 & 0.094 & 0.220 & 0.95 & 0.94\\
 & 1.5 & 0.90 & 0.009 & 0.128 & 0.93 & 1.05 & 0.049 & 0.203 & 0.95 & 0.99\\
 & 4 & 0.25 & 0.010 & 0.068 & 0.90 & 0.96 & 0.896 & 1.680 & 0.95 & 0.78\\
 & 4 & 0.50 & -0.013 & 0.055 & 0.97 & 0.88 & 0.998 & 1.504 & 0.97 & 0.59\\
 & 4 & 0.90 & -0.022 & 0.100 & 0.93 & 0.96 & 0.971 & 1.457 & 1.00 & 0.58\\
\bottomrule\end{tabular}\end{table}

\emph{Regression coefficients.} The coefficients are recovered accurately in every design. The bias of $\hat\beta_1$ never exceeds 0.04 in absolute value, against RMSEs of 0.11--0.40 at $n=300$ and 0.06--0.17 at $n=1{,}000$; averaged over $r$, the RMSE falls from 0.14 to 0.08 at $\tau=0.5$ and from 0.28 to 0.14 at $\tau=0.9$, close to the $\sqrt n$ rate ($\sqrt{1000/300}=1.83$). Upper quantiles are harder to estimate because fewer observations lie near them, and the heavier the body (smaller $r$), the larger the RMSE at $\tau=0.9$. The credible intervals cover the true slope in 90--100\% of replications, which is within Monte Carlo error (standard error 0.035) of the nominal 95\%, and the ratio E/P of the empirical standard deviation of the posterior mean to the average posterior standard deviation has median 0.97 (range 0.75--1.08 over the 18 cells, within Monte Carlo variation for 40 replications), so that the posterior spread matches the sampling variability. The intercept and $\sigma$ behave in the same way, with biases at $n=1{,}000$ below 0.02 and 0.05 respectively.

\emph{Power parameter.} The power is recovered well when it is small or moderate. For $r=0.75$ the bias of the posterior mean is 0.08 at $n=300$ and 0.015 at $n=1{,}000$; for $r=1.5$ it falls from 0.58 to 0.07, the larger value at $n=300$ reflecting a right-skewed posterior. Coverage is 90--100\% and E/P between 0.67 and 1.14. For $r=4$ the posterior is wide and right-skewed even at $n=1{,}000$ (median length of the 95\% interval 7.4): the posterior mean overestimates $r$ by about one unit and the posterior median by 0.5, while the intervals still cover the true value in 95--100\% of replications and are, if anything, conservative (E/P 0.49--0.78). This weak identification of large powers is predicted by Theorem~\ref{thm:limit}: as $r$ grows the PSMALD density converges to PSMALD$_\infty$, so the likelihood flattens in $r$, and $r=4$ can be distinguished from $r=8$ only through small differences in the shape of the body.

\begin{figure}[t]\centering
\includegraphics[width=0.95\textwidth]{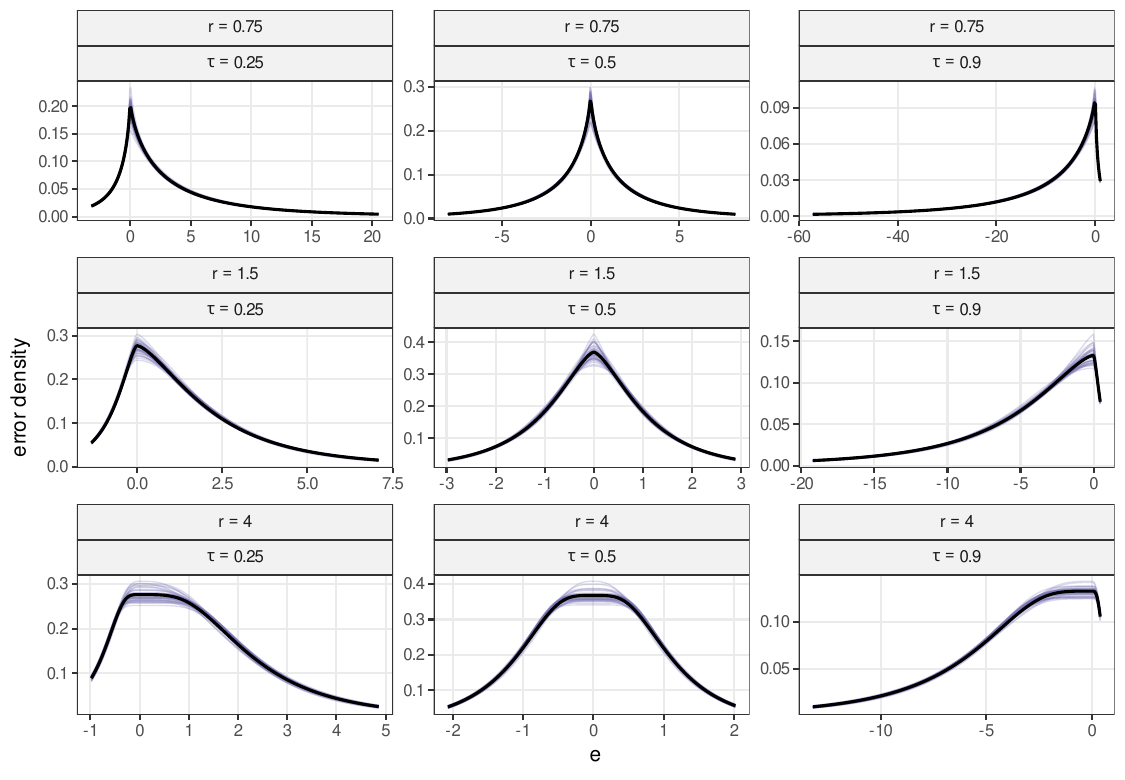}
\caption{Simulation study 1 ($n=1{,}000$): true PSMALD error density (black) and the 40 densities implied by the posterior means of $(\sigma,r)$ in the 40 replications (purple), by true power $r$ (rows) and quantile level $\tau$ (columns).}\label{fig:sim_density}
\end{figure}

\emph{The fitted error density.} What matters for density forecasts is not $r$ alone but the density it implies. Figure~\ref{fig:sim_density} overlays, for $n=1{,}000$, the true PSMALD error density with the 40 densities implied by the posterior means of $(\sigma,r)$. For $r=0.75$ and $r=1.5$ the fitted densities are almost indistinguishable from the truth; for $r=4$ they vary somewhat in the flatness of the top but agree closely in the shoulders and tails, where $a_\lambda$ fixes the decay exponent and $r$ and $\sigma$ affect the tail constants (Theorem~\ref{thm:tail}). In this experiment, the fitted densities remain close to the generating density even when $r$ is estimated imprecisely.

\subsection{Comparison under misspecification and validation of Proposition~\ref{prop:pseudo}}\label{sec:sim_misspec}
\begin{figure}[t]\centering
\includegraphics[width=\textwidth]{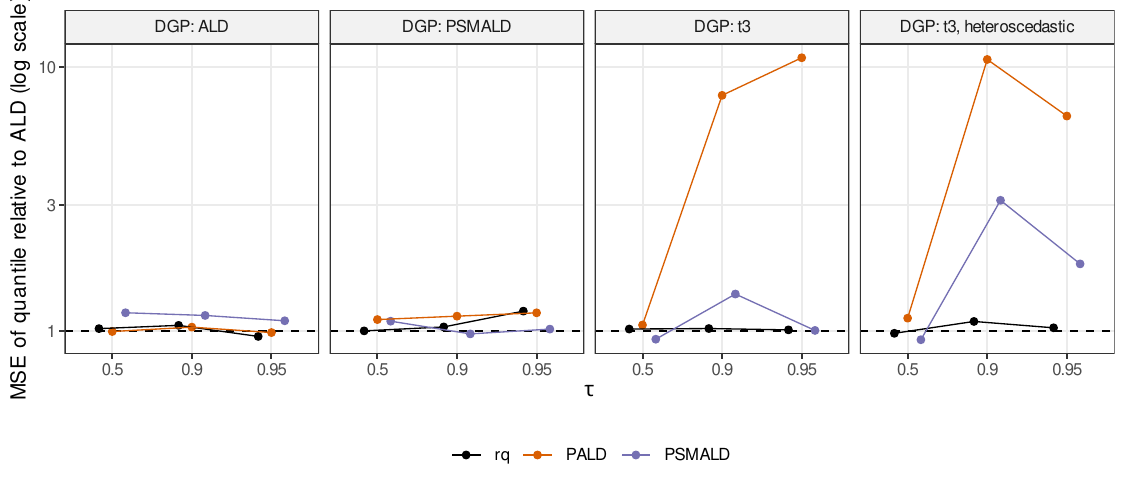}
\caption{Simulation study 2: mean squared error of the estimated conditional quantile on the test points, relative to the ALD (log scale; values below 1 favour the model), by data-generating process and $\tau$.}\label{fig:sim_misspec}
\end{figure}

The second study fits the ALD, PALD and PSMALD, together with the exact check-loss estimator (rq), to data from four processes: ALD errors; PSMALD errors with $r=1.5$ and $a_\lambda=4$; Student-$t_3$ errors; and Student-$t_3$ errors with heteroscedastic scale $s(x)=1+0.75x$. The errors are shifted so that their $\tau$-quantile is zero. The first two designs are correctly specified for the ALD and for PSMALD respectively, and the last two for none of the models, which provide two misspecification scenarios. We use $n=1{,}000$ training and $1{,}000$ test observations, $\tau\in\{0.5,0.9,0.95\}$ and $R=30$ replications. Figure~\ref{fig:sim_misspec} reports the mean squared error of $\hat Q(\tau\mid x)$ over the test points relative to the ALD, and Supplementary Table~\ref{S-tab:sim_misspec_full} gives the bias and RMSE of each coefficient, the coverage of the credible intervals, the test check loss and the test coverage.

\emph{Correct specification.} Under ALD errors, PALD has quantile MSE within 4\% of the ALD. Under PSMALD errors, the PSMALD fit improves on the ALD at $\tau=0.90$ (relative MSE 0.98) and gives nearly the same error at $\tau=0.95$ (1.02). At the median its relative MSE is 1.09. The nominal 95\% credible intervals for the quantile cover in 93--94\% of replications on average over the three evaluation points. PALD has 11--18\% higher quantile MSE than the ALD in the PSMALD designs. Under ALD errors, PSMALD has 10--18\% higher quantile MSE than the ALD.

\emph{Heavy-tailed errors.} The $t_3$ designs show why the scale-mixture layer of the framework matters. At $\tau=0.5$ the errors are symmetric, every model targets the median, and all perform alike; PSMALD is the most accurate of all at this level in both $t_3$ designs (relative MSE 0.93 and 0.93). At $\tau=0.9$ and $0.95$ the exponential tails of PALD are too light: its relative MSE rises to 7.8 and 10.9, its intercept is biased by $-0.31$ and $-0.76$, and its credible intervals cover the true quantile in only 19\% and 7\% of replications. PSMALD, whose polynomial tails have a survival exponent of 4, compared with 3 for the $t_3$ law, reduces this loss by a factor of 5.6 and 11 to a relative MSE of 1.39 and 1.01, and in the homoscedastic design it is as accurate as the ALD at $\tau=0.95$. Under heteroscedasticity the pseudo-true slopes of both flexible models are shifted (Section~\ref{sec:pseudo}); PSMALD keeps the loss to a relative MSE of 1.8--3.1, against 6.5--10.7 for PALD, while the ALD and rq remain accurate. The ALD point estimates remain competitive across the designs, and the ALD credible intervals are close to nominal under correct specification (coverage 90--96\%); under $t_3$ errors at $\tau\ge0.9$ they cover in only 53--82\% of replications, because the ALD quasi-posterior is too narrow under misspecification \citep{yang2016}.

\begin{figure}[t]\centering
\includegraphics[width=\textwidth]{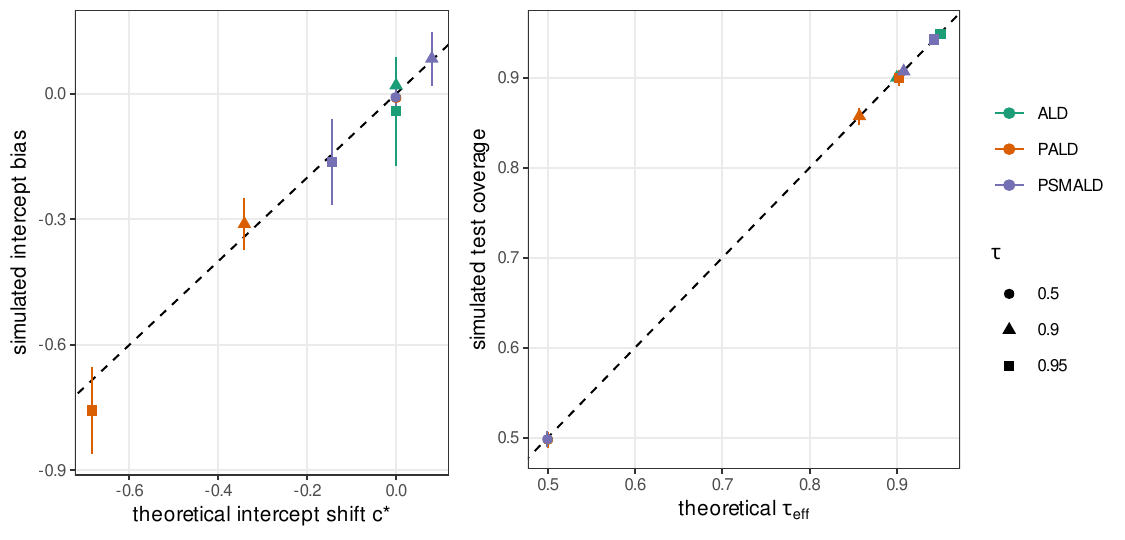}
\caption{Simulation study 2, Student-$t_3$ errors: validation of Proposition~\ref{prop:pseudo}. Left: Monte Carlo mean of the intercept error ($\pm2$ standard errors) against the theoretical pseudo-true shift $c^*$. Right: empirical test coverage against the theoretical effective level $\tau_{\rm eff}$. Dashed line: identity.}\label{fig:sim_pseudo}
\end{figure}

\emph{Validation of Proposition~\ref{prop:pseudo}.} Figure~\ref{fig:sim_pseudo} confronts these results with the theory. The left panel plots the Monte Carlo mean of the intercept error in the homoscedastic $t_3$ design against the theoretical shift $c^*$, computed by \texttt{pseudo\_true} from the $t_3$ law alone; the right panel plots the empirical test coverage against the theoretical effective level $\tau_{\rm eff}$. The estimates generally track the numerical projections. At $\tau=0.9$, PALD has $c^*=-0.34$ and $\tau_{\rm eff}=0.857$ against a simulated bias of $-0.31$ and coverage of $0.857$, and at $\tau=0.95$ PSMALD has $c^*=-0.14$ and $\tau_{\rm eff}=0.943$ against $-0.16$ and $0.943$. These comparisons show how the numerical population projections help interpret the empirical coverage in this homoscedastic design.

\emph{Summary.} The Bayesian procedure recovers the parameters of the proposed model with empirical coverage near the nominal level in most designs for the regression coefficients; the relative accuracy of the correctly specified model varies across quantile levels; PSMALD reduces upper-quantile error relative to PALD in the Student-$t_3$ designs, while the ALD remains competitive; and Proposition~\ref{prop:pseudo} gives a population explanation of these differences.

%=====================================================================
\section{Case studies: forecasting daily tail risk}\label{sec:case}
%=====================================================================
\subsection{Data and design}\label{sec:case_data}
We apply the framework to daily percentage losses $L_t=-100\log(P_t/P_{t-1})$ of three emerging-market series for which we compare tail forecasts: (i) the Nifty Bank index of the National Stock Exchange of India; (ii) the Brazilian Ibovespa (BOVESPA), an equity market exposed to commodity, currency and political shocks; and the (iii) Indian rupee against the US dollar, a managed currency whose losses are rupee depreciations. The three series cover two asset classes and two economies, and differ in the size and frequency of large losses. For the rupee, $P_t$ is the reciprocal of the quoted rupees per US dollar. Prices are Yahoo Finance daily closes from January 2000, or the first available date, to December 2025, frozen in the R package. Table~\ref{tab:data} summarises the samples; the largest daily losses, between 6.1\% (rupee) and 18.3\% (Nifty Bank), are many standard deviations from the mean.

\begin{table}[t]\centering\small
\caption{The three case-study series: daily percentage losses $L_t=-100\log(P_t/P_{t-1})$ from Yahoo Finance daily closes (downloaded on 23 November 2026 and frozen in the R package). The first 80\% of each series is the training sample and the last 20\% the test sample; INR/USD is the US-dollar value of one rupee, so a loss is a depreciation of the rupee. $\hat\xi$: generalised Pareto shape of the 5\% largest losses with 95\% interval; the last column is the $p$-value of the likelihood-ratio test of an exponential tail, $\xi=0$, using the nominal chi-squared reference distribution.}
\label{tab:data}\setlength{\tabcolsep}{4pt}
\resizebox{\textwidth}{!}{\begin{tabular}{lll rr rr c c}\toprule
Series & Ticker & Sample & $n_{\rm train}$ & $n_{\rm test}$ & Mean & SD & GPD shape $\hat\xi$ (95\% CI) & $p$-value, $\xi=0$\\ \midrule
Bank Nifty & \texttt{\^{}NSEBANK} & 10/2007--12/2025 & 3,584 & 896 & -0.043 & 1.73 & 0.17 (0.02, 0.33) & 0.005\\
BOVESPA & \texttt{\^{}BVSP} & 02/2000--12/2025 & 5,136 & 1,284 & -0.035 & 1.70 & 0.26 (0.13, 0.39) & $1.9\times10^{-7}$\\
INR/USD & \texttt{INR=X}$^{-1}$ & 12/2003--12/2025 & 4,564 & 1,142 & 0.012 & 0.47 & 0.29 (0.14, 0.44) & $4.5\times10^{-8}$\\
\bottomrule\end{tabular}}\end{table}

\begin{figure}[!htbp]\centering
\includegraphics[width=\textwidth]{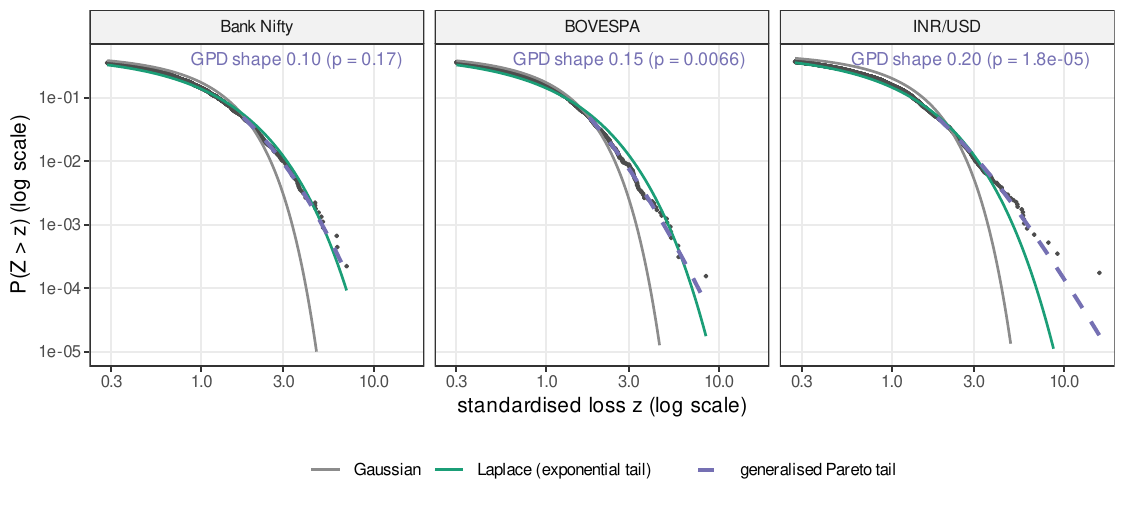}
\caption{Right tails of the volatility-standardised losses $z_t=L_t/\mathrm{Volatility}_{20,t-1}$: empirical survival function $P(Z>z)$ on log--log axes (points), with the Gaussian and Laplace (exponential-tail, ALD-type) tails fitted to the same data, and the generalised Pareto tail fitted to the 5\% largest values (dashed), with its shape $\hat\xi$ and the $p$-value of the test of an exponential tail ($\xi=0$). An exponential tail bends downwards on this scale; a polynomial tail, as in PSMALD, is close to a straight line.}\label{fig:tails}
\end{figure}

\emph{Tail and central-shape diagnostics.} Table~\ref{tab:data} and Figure~\ref{fig:tails} describe the raw losses $L_t$ and the volatility-standardised losses $z_t=L_t/\mathrm{Volatility}_{20,t-1}$. When comparing the two, note that $z_t$ adjusts for lagged volatility only. The excess kurtosis of $L_t$ is 7.0 for BOVESPA, 9.6 for Nifty Bank and 21.0 for the rupee, above the Laplace benchmark of 3. Generalised Pareto fits to the largest 5\% give positive shape estimates for all three series, ranging from 0.17 to 0.29 for raw losses and 0.10 to 0.20 after standardisation. A threshold likelihood-ratio test of zero shape rejects for all three raw series at the nominal 1\% level. After standardisation, the nominal $p$-values are 0.169 for Nifty Bank, 0.007 for BOVESPA and below $10^{-4}$ for the rupee. Figure~\ref{fig:tails} shows the corresponding tail fits. The Laplace benchmark assigns less probability to the largest losses than the empirical survival curve, particularly for the rupee. These threshold comparisons are descriptive and their nominal reference distributions do not account for serial dependence. They also do not identify a tail law conditional on the full set of regression predictors. The distributions differ near the centre as well. For $z_t$, the fraction of observations within 0.25 standard deviations of the median is 0.22--0.28, between the Gaussian value of 0.20 and the Laplace value of 0.30.

These diagnostics motivate three comparisons. First, the estimated power of PALD allows departures from the Laplace shape, which may improve check loss at moderate levels. Second, PALD uses the same power to determine the central shape and tail decay. Values below one slow the tail decay relative to the ALD, but the tails remain stretched exponential. Section~\ref{sec:pseudo} illustrates how this restriction can affect calibration under heavy-tailed misspecification. Third, PSMALD fixes a polynomial survival exponent through $a_\lambda=4$, corresponding to $\xi=0.25$ in the working distribution. The power can then change the kernel shape while preserving that exponent. The holdout comparison examines how these features affect quantile and density forecasts in these three series.

For each series and each $\tau\in\{0.90,0.95,0.97,0.99\}$, the levels used in value-at-risk reporting, we model the conditional $\tau$-quantile of $L_t$ as a linear function of three predictors known at the end of day $t-1$: the previous loss $L_{t-1}$, which captures short-term persistence and the leverage effect; the previous squared return $L_{t-1}^2$, an ARCH-type shock term; and the standard deviation of the 20 previous returns, a measure of recent volatility. The first 80\% of each series is used for estimation and the last 20\%, between 896 and 1,284 days, for out-of-sample evaluation. The training samples include the 2008 financial crisis and the March 2020 crash. The test samples begin on 17 May 2022 for Nifty Bank, 4 November 2020 for BOVESPA and 10 August 2021 for INR/USD. Their loss standard deviations are lower than in the corresponding training samples. We fit PALD and PSMALD and compare them with the ALD, the standard working likelihood, and with the exact check-loss estimator (rq), in 12 series--level combinations.

\subsection{Posterior estimates}\label{sec:case_post}
Table~\ref{tab:post} reports the PSMALD posterior at $\tau=0.95$ and $0.99$. In every series the 20-day volatility is the dominant predictor, with posterior means between 0.61 and 1.04 and credible intervals far from zero: holding the other predictors fixed, a one-point increase in recent volatility is associated with a 0.7--1.0 percentage-point increase in the fitted 99\% VaR. The lagged loss has a positive effect for the two equity indices (0.07--0.19), consistent with short-term persistence after losses, but a negative effect for the rupee ($-0.04$ and $-0.11$), consistent with mean reversion of a managed exchange rate, for which the lagged squared return carries the shock instead (0.11--0.12). The power $r$ is large, 17--27, and $\sigma$ is well determined. Supplementary Table~\ref{S-tab:post_psmal} reports the posterior at all four levels.

\begin{table}[!htbp]\centering\footnotesize
\caption{Posterior estimates of the PSMALD quantile regression of the daily loss on the lagged loss, lagged squared return and 20-day volatility, on the original (percentage) scale: posterior mean and 95\% credible interval. The last column gives the ALD posterior mean of the volatility coefficient for comparison.}
\label{tab:post}\setlength{\tabcolsep}{3pt}\resizebox{\textwidth}{!}{\begin{tabular}{ll cccccc c}\toprule
Series & $\tau$ & Intercept & $L_{t-1}$ & $L_{t-1}^2$ & Volatility$_{20}$ & $\sigma$ & $r$ & ALD: Volatility$_{20}$\\ \midrule
Bank Nifty & 0.95 & \begin{tabular}[c]{@{}c@{}}1.142\\[-2pt]{\scriptsize(1.038, 1.237)}\end{tabular} & \begin{tabular}[c]{@{}c@{}}0.191\\[-2pt]{\scriptsize(0.167, 0.213)}\end{tabular} & \begin{tabular}[c]{@{}c@{}}-0.017\\[-2pt]{\scriptsize(-0.020, -0.014)}\end{tabular} & \begin{tabular}[c]{@{}c@{}}0.894\\[-2pt]{\scriptsize(0.846, 0.955)}\end{tabular} & \begin{tabular}[c]{@{}c@{}}0.317\\[-2pt]{\scriptsize(0.307, 0.326)}\end{tabular} & \begin{tabular}[c]{@{}c@{}}17.1\\[-2pt]{\scriptsize(11.3, 25.4)}\end{tabular} & 1.389\\
 & 0.99 & \begin{tabular}[c]{@{}c@{}}1.895\\[-2pt]{\scriptsize(1.779, 2.016)}\end{tabular} & \begin{tabular}[c]{@{}c@{}}0.086\\[-2pt]{\scriptsize(0.059, 0.120)}\end{tabular} & \begin{tabular}[c]{@{}c@{}}-0.016\\[-2pt]{\scriptsize(-0.019, -0.011)}\end{tabular} & \begin{tabular}[c]{@{}c@{}}1.035\\[-2pt]{\scriptsize(0.957, 1.098)}\end{tabular} & \begin{tabular}[c]{@{}c@{}}0.080\\[-2pt]{\scriptsize(0.078, 0.083)}\end{tabular} & \begin{tabular}[c]{@{}c@{}}22.5\\[-2pt]{\scriptsize(15.5, 32.5)}\end{tabular} & 2.122\\
\midrule
BOVESPA & 0.95 & \begin{tabular}[c]{@{}c@{}}1.372\\[-2pt]{\scriptsize(1.259, 1.504)}\end{tabular} & \begin{tabular}[c]{@{}c@{}}0.071\\[-2pt]{\scriptsize(0.047, 0.096)}\end{tabular} & \begin{tabular}[c]{@{}c@{}}-0.018\\[-2pt]{\scriptsize(-0.024, -0.011)}\end{tabular} & \begin{tabular}[c]{@{}c@{}}0.867\\[-2pt]{\scriptsize(0.789, 0.933)}\end{tabular} & \begin{tabular}[c]{@{}c@{}}0.341\\[-2pt]{\scriptsize(0.333, 0.350)}\end{tabular} & \begin{tabular}[c]{@{}c@{}}22.4\\[-2pt]{\scriptsize(15.6, 31.2)}\end{tabular} & 1.211\\
 & 0.99 & \begin{tabular}[c]{@{}c@{}}2.118\\[-2pt]{\scriptsize(2.043, 2.181)}\end{tabular} & \begin{tabular}[c]{@{}c@{}}0.192\\[-2pt]{\scriptsize(0.173, 0.206)}\end{tabular} & \begin{tabular}[c]{@{}c@{}}-0.018\\[-2pt]{\scriptsize(-0.021, -0.015)}\end{tabular} & \begin{tabular}[c]{@{}c@{}}1.002\\[-2pt]{\scriptsize(0.973, 1.041)}\end{tabular} & \begin{tabular}[c]{@{}c@{}}0.084\\[-2pt]{\scriptsize(0.083, 0.086)}\end{tabular} & \begin{tabular}[c]{@{}c@{}}26.6\\[-2pt]{\scriptsize(18.5, 36.4)}\end{tabular} & 1.711\\
\midrule
INR/USD & 0.95 & \begin{tabular}[c]{@{}c@{}}0.415\\[-2pt]{\scriptsize(0.374, 0.457)}\end{tabular} & \begin{tabular}[c]{@{}c@{}}-0.040\\[-2pt]{\scriptsize(-0.068, -0.015)}\end{tabular} & \begin{tabular}[c]{@{}c@{}}0.119\\[-2pt]{\scriptsize(0.111, 0.133)}\end{tabular} & \begin{tabular}[c]{@{}c@{}}0.611\\[-2pt]{\scriptsize(0.525, 0.697)}\end{tabular} & \begin{tabular}[c]{@{}c@{}}0.087\\[-2pt]{\scriptsize(0.085, 0.090)}\end{tabular} & \begin{tabular}[c]{@{}c@{}}19.8\\[-2pt]{\scriptsize(13.3, 29.4)}\end{tabular} & 1.192\\
 & 0.99 & \begin{tabular}[c]{@{}c@{}}0.631\\[-2pt]{\scriptsize(0.583, 0.706)}\end{tabular} & \begin{tabular}[c]{@{}c@{}}-0.106\\[-2pt]{\scriptsize(-0.138, -0.051)}\end{tabular} & \begin{tabular}[c]{@{}c@{}}0.114\\[-2pt]{\scriptsize(0.096, 0.144)}\end{tabular} & \begin{tabular}[c]{@{}c@{}}0.696\\[-2pt]{\scriptsize(0.569, 0.787)}\end{tabular} & \begin{tabular}[c]{@{}c@{}}0.022\\[-2pt]{\scriptsize(0.021, 0.023)}\end{tabular} & \begin{tabular}[c]{@{}c@{}}24.6\\[-2pt]{\scriptsize(17.6, 34.1)}\end{tabular} & 1.894\\
\bottomrule\end{tabular}}\end{table}

The ALD gives a larger volatility coefficient and a smaller intercept in all six comparisons in Table~\ref{tab:post}. The volatility coefficient is about 1.4 to 2.7 times as large under the ALD and reaches 2.12. PSMALD assigns large fitted scales to some observations from 2008 and 2020, as discussed in Section~\ref{sec:case_density}. This pattern is consistent with the different way the two likelihoods handle large residuals. The fitted coefficient differences also reflect other differences between the two likelihoods and cannot be attributed to particular observations from this comparison alone.

\subsection{Where the proposed models improve on the ALD}\label{sec:case_gains} 
\begin{figure}[t]\centering 
\includegraphics[width=\textwidth]{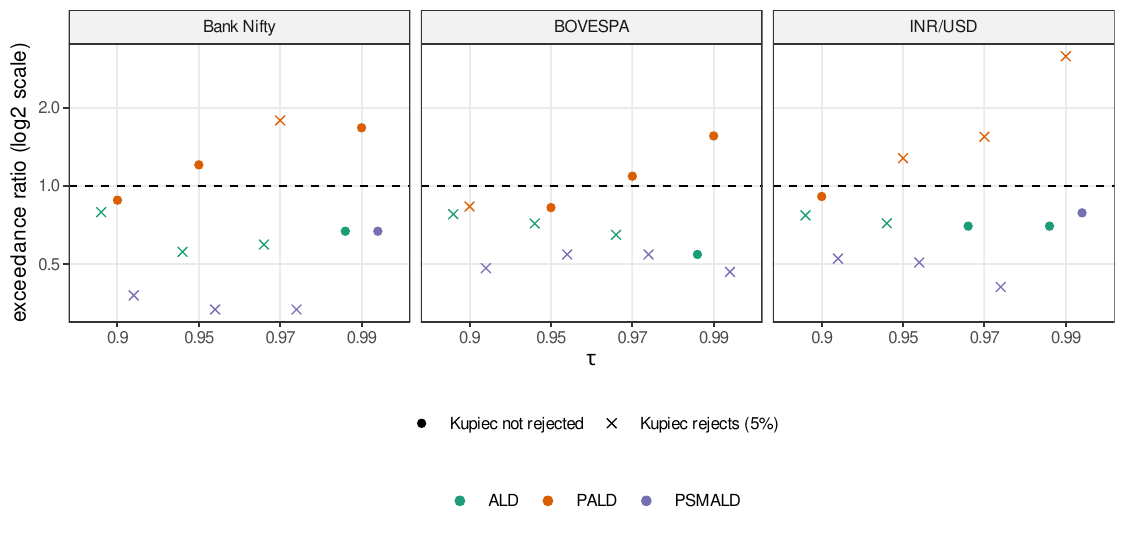}
\caption{Test-sample exceedance ratio of the ALD, PALD and PSMALD by series and $\tau$ (log scale; 1 is perfect calibration). Crosses mark rejection by the Kupiec test at the 5\% level. The 99\% PALD points for Nifty Bank and INR/USD are provisional.}\label{fig:case_exc}
\end{figure}

Table~\ref{tab:summary} summarises the comparison over the 12 series--level combinations; Table~\ref{tab:checkloss} gives the check losses with Diebold--Mariano tests, Figure~\ref{fig:case_exc} the exceedance ratios, and Supplementary Table~\ref{S-tab:backtests_full} the complete backtests. In the archived comparisons, PALD or PSMALD has a lower test check loss than the ALD in 11 of the 12 combinations and an exceedance ratio closer to one in 9, and the ALD has the lowest check loss of the three models in only one combination. Five improvements meet the nominal 5\% directional Diebold--Mariano threshold, four for PALD and one for PSMALD, while the archived tests do not detect a significant gain for the ALD over PALD.

\begin{table}[t]\centering\small
\caption{Descriptive rankings of the archived fits, including the provisional Nifty Bank and INR/USD PALD fits at 0.99. Summary of the case studies: number of the 12 series--level combinations (three series, four levels) in which each working likelihood is best on each criterion, and passes both the Kupiec and DQ tests at the 5\% level. Diebold--Mariano comparisons are in Table~\ref{tab:checkloss} and Supplementary Table~\ref{S-tab:dm_full}.}
\label{tab:summary}\begin{tabular}{l ccc}\toprule
Criterion & ALD & PALD & PSMALD\\ \midrule
\multicolumn{4}{l}{\textit{Density forecasts}}\\
\quad Best test log score & 0 & 0 & 12\\
\quad Best training LOOIC & 0 & 0 & 12\\
\multicolumn{4}{l}{\textit{Quantile forecasts}}\\
\quad Lowest test check loss & 1 & 9 & 2\\
\quad Smallest absolute log exceedance ratio & 2 & 8 & 2\\
\quad Passes Kupiec and DQ tests & 4 & 5 & 2\\
\bottomrule\end{tabular}\end{table}

\emph{PALD at the lower quantile levels.} PALD has lower test check loss than the ALD in 10 of the 12 archived comparisons. It has the lowest loss among the three Bayesian fits in nine, including the lower three levels for both Nifty Bank and BOVESPA. The count includes the provisional 99\% result for Nifty Bank, where the archived loss is 4.1\% lower than under the ALD. The improvements at $\tau=0.90$ for Nifty Bank and at $\tau=0.90$, $0.95$ and $0.97$ for BOVESPA meet the nominal 5\% directional Diebold--Mariano threshold. PALD has the smallest absolute log exceedance ratio in eight of the archived comparisons and passes both the Kupiec and DQ tests in five, compared with four for the ALD. These summaries describe this particular holdout and include the two provisional 99\% PALD estimates.

\begin{figure}[!htbp]\centering
\includegraphics[width=\textwidth]{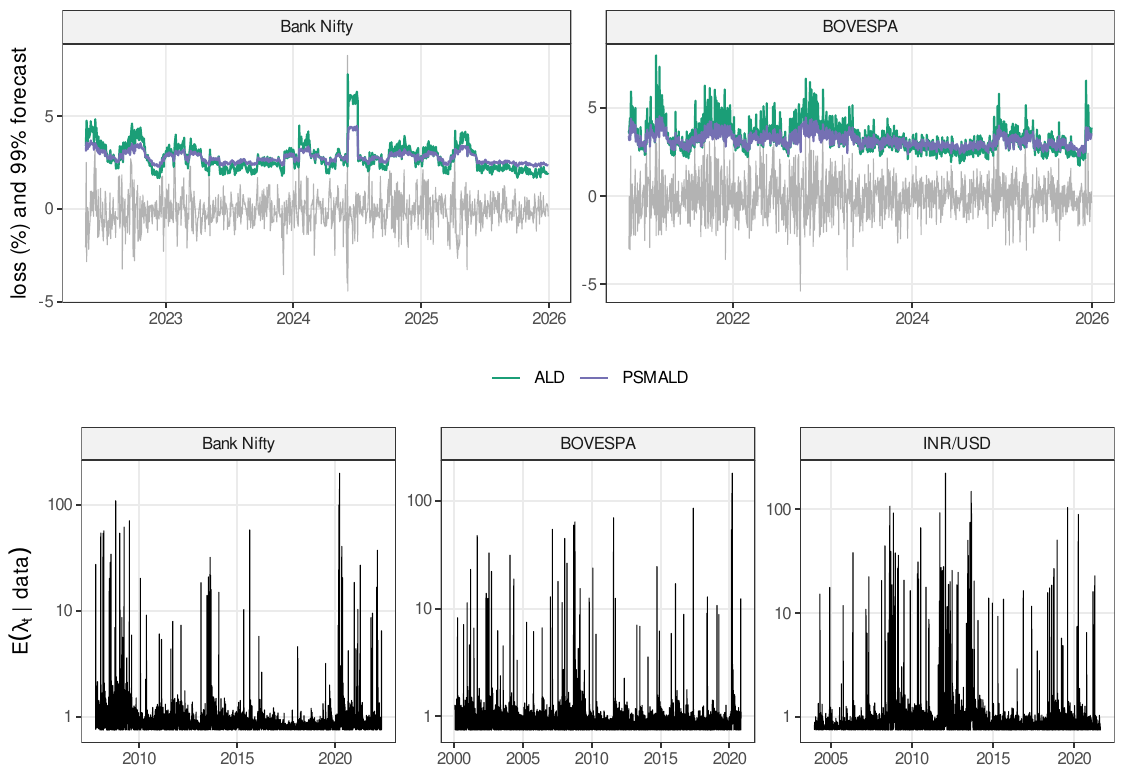}
\caption{The 99\% level. Top: test-sample losses (grey) of Nifty Bank and BOVESPA with the ALD and PSMALD quantile forecasts. Bottom: posterior mean of the latent scale $\lambda_t$ (PSMALD, log scale) over the training sample of each series.}\label{fig:case_lambda}
\end{figure}

\emph{PSMALD at the 99\% level.} At $\tau=0.99$, PSMALD reduces check loss relative to the ALD by 3.3\% for Nifty Bank, 3.4\% for BOVESPA and 1.9\% for the rupee. The corresponding two-sided Diebold--Mariano $p$-values are 0.127, 0.006 and 0.430. The PSMALD forecasts pass both the Kupiec and the DQ tests for Nifty Bank and the rupee. For the rupee, the PSMALD exceedance ratio is 0.79, compared with 0.70 under the ALD and 3.15 in the provisional PALD fit. Figure~\ref{fig:case_lambda} illustrates the trajectories of the ALD and PSMALD forecasts. The pairwise comparisons in Supplementary Table~\ref{S-tab:dm_full} generally favour PALD over PSMALD at 90\%. At 99\% the statistics favour PSMALD for BOVESPA and the rupee, although the latter comparison involves a poorly mixing PALD fit.

The comparison also shows the limits of each member, and these follow the same logic. Loss volatility falls in the test period for each series. Over-coverage is common in the holdouts: the ALD exceedance ratio is below one in all 12 combinations, and all 8 of its Kupiec rejections are due to over-coverage. PSMALD, whose tails are the heaviest, amplifies this conservatism at $\tau$ between 0.90 and 0.97 (exceedance ratios 0.33--0.55), which is where its check loss exceeds that of the ALD; at these levels PALD generally gives lower check loss than PSMALD in this sample. The Bayesian ALD and rq produce similar average check losses.

\begin{table}[t]\centering\footnotesize
\caption{Test-sample mean check loss of ALD and percentage change of the other working likelihoods relative to ALD (negative is better). The Nifty Bank and INR/USD PALD entries at 0.99 are provisional. Diebold--Mariano tests use Newey--West variances. $^{+}$ denotes lower loss at the 5\% level in a one-sided test, and $^{-}$ denotes higher loss in a two-sided test at the same nominal level. Bold: lowest loss in each row.}
\label{tab:checkloss}\setlength{\tabcolsep}{5pt}
\begin{tabular}{ll c cc}\toprule
Asset & $\tau$ & ALD & PALD & PSMALD\\ \midrule
Bank Nifty & 0.90 & 0.1737 & \textbf{-1.3$^{+}$} & +8.2$^{-}$\\
 & 0.95 & 0.1108 & \textbf{-2.3} & +4.9$^{-}$\\
 & 0.97 & 0.0797 & \textbf{-2.1} & +1.6\\
 & 0.99 & 0.0390 & \textbf{-4.1} & -3.3\\
\midrule
BOVESPA & 0.90 & 0.2089 & \textbf{-0.4$^{+}$} & +6.3$^{-}$\\
 & 0.95 & 0.1293 & \textbf{-1.3$^{+}$} & +2.6$^{-}$\\
 & 0.97 & 0.0877 & \textbf{-2.3$^{+}$} & +2.2\\
 & 0.99 & 0.0377 & -1.6 & \textbf{-3.4$^{+}$}\\
\midrule
INR/USD & 0.90 & 0.0513 & \textbf{-0.9} & +7.5$^{-}$\\
 & 0.95 & 0.0328 & \textbf{-0.7} & +4.1$^{-}$\\
 & 0.97 & \textbf{0.0232} & +1.8 & +3.3\\
 & 0.99 & 0.0111 & +9.5 & \textbf{-1.9}\\
\bottomrule\end{tabular}\end{table}

\subsection{Density forecasts, the power parameter and the latent scales}\label{sec:case_density}
For density forecasts, the comparison favours PSMALD (Table~\ref{tab:density}). PSMALD has the highest test log score in all 12 combinations, exceeding the ALD by 0.08 to 0.45 nats per day (median 0.20), and PALD also improves on the ALD in all 12. The training-sample LOOIC selects PSMALD in all 12 combinations, with differences from the ALD of hundreds to thousands of units (Supplementary Table~\ref{S-tab:ic_full}). The scores measure the quality of the full predictive distribution and need not agree with the check-loss rankings.

\begin{table}[!htbp]\centering\footnotesize
\caption{Density fit. Left: mean test-sample log predictive density $n_{\rm test}^{-1}\sum_t\log\hat p(L_t)$ (higher is better; bold: best). Right: mean of the stored posterior subsample for the power parameter $r$ for PALD and PSMALD. For the provisional Nifty Bank and INR/USD PALD fits at 0.99, see Supplementary Table~\ref{S-tab:convergence}.}
\label{tab:density}\setlength{\tabcolsep}{4pt}
\begin{tabular}{ll ccc cc}\toprule
 & & \multicolumn{3}{c}{Test log score} & \multicolumn{2}{c}{Posterior mean of $r$}\\ \cmidrule(lr){3-5}\cmidrule(lr){6-7}
Asset & $\tau$ & ALD & PALD & PSMALD & PALD & PSMALD\\ \midrule
Bank Nifty & 0.90 & -1.788 & -1.757 & \textbf{-1.693} & 0.81 & 14.8\\
 & 0.95 & -1.968 & -1.892 & \textbf{-1.774} & 0.70 & 17.0\\
 & 0.97 & -2.112 & -1.977 & \textbf{-1.845} & 0.59 & 18.9\\
 & 0.99 & -2.428 & -2.195 & \textbf{-1.989} & 0.53 & 22.4\\
\midrule
BOVESPA & 0.90 & -1.914 & -1.911 & \textbf{-1.829} & 0.95 & 20.1\\
 & 0.95 & -2.068 & -2.060 & \textbf{-1.911} & 0.88 & 22.4\\
 & 0.97 & -2.169 & -2.148 & \textbf{-1.970} & 0.80 & 24.2\\
 & 0.99 & -2.392 & -2.350 & \textbf{-2.078} & 0.69 & 26.4\\
\midrule
INR/USD & 0.90 & -0.545 & -0.512 & \textbf{-0.432} & 0.79 & 16.6\\
 & 0.95 & -0.728 & -0.656 & \textbf{-0.524} & 0.70 & 19.8\\
 & 0.97 & -0.868 & -0.752 & \textbf{-0.611} & 0.61 & 22.0\\
 & 0.99 & -1.173 & -0.956 & \textbf{-0.724} & 0.49 & 24.6\\
\bottomrule\end{tabular}\end{table}

Table~\ref{tab:density} shows contrasting patterns in the estimated powers. For PALD, the reported means decline from 0.79--0.95 at $\tau=0.90$ to 0.48--0.69 at $\tau=0.99$. A smaller power slows the decay of the PALD density relative to the ALD and also changes its central shape. For PSMALD, the reported means increase from 15--20 at $\tau=0.90$ to 22--27 at $\tau=0.99$. A large power gives a rounded, nearly flat peak and can bring the fitted density close to the uniform-mixture limit of Theorem~\ref{thm:limit}. This interpretation is consistent with the broad intervals in Tables~\ref{tab:post} and \ref{S-tab:post_psmal}, and with the weak identification of large powers in the simulation study. The two 99\% PALD summaries flagged below are provisional.

The latent scales provide an interpretable by-product that the ALD cannot offer. The posterior mean of $\lambda_t$ is close to one on ordinary days and rises sharply on crisis days (Figure~\ref{fig:case_lambda}). For Nifty Bank the largest value is on 23 March 2020 ($\E(\lambda_t\mid\bm y)=200$), during the COVID-19 market sell-off in India, followed by 24 October 2008 (110) during the global financial crisis; for BOVESPA the largest values are on 16 and 12 March 2020 (180 and 135), and a prominent one on 18 May 2017 (85), the day of the political-scandal crash in Brazil; for the rupee the second-largest value (149) is on 27 August 2013, at the height of the ``taper tantrum''. These days are identified from the losses and the three predictors alone, and they have large residual scales under the fitted likelihood.

\subsection{Synthesis}\label{sec:case_summary}
The holdout comparison gives different rankings for quantile and density forecasts. For the quantile at the levels used in risk management, PALD is the fit with the lowest check loss and the strongest calibration by the reported summaries in most archived combinations, with the qualifications about the two 99\% PALD estimates above. At the 99\% level, the highest level considered, PSMALD improves on the ALD in every series. PSMALD also has the highest held-out mean log predictive density in all 12 comparisons. Its fitted scales describe observations with large residuals in the training period. Because PALD, PSMALD and the ALD belong to one nested family and are fitted with the same software, the member can be chosen per series and per level, guided by the log score for the distribution, by the backtests for the quantile, and by the effective level $\tau_{\rm eff}$ of Proposition~\ref{prop:pseudo} for what each member is expected to deliver.

%=====================================================================
\section{Conclusion}\label{sec:conclusion}
%=====================================================================
We have proposed a Bayesian quantile regression framework built on a nested family of quantile-preserving working likelihoods. Raising the check loss to a power gives the power asymmetric Laplace distribution (PALD), which frees the shape of the body; mixing its scale over observation-specific inverse-gamma variables gives the PSMALD distribution, which adds polynomial tails and lets each observation carry its own scale. The ALD is the member with both features switched off, so the framework contains the standard model and lets the data decide how far to move away from it.

We derived the distribution and quantile functions and moments of the PALD law, the marginal PSMALD density, its tail exponent and the closed-form limit as $r\to\infty$. The last result provides a reference point for interpreting large fitted powers. For a homoscedastic location model, the misspecification result gives the effective quantile level associated with the population fit when a reference error law is available. The simulations compare this calculation with finite-sample coverage. For computation, two representations of the marginal density permit numerical integration of the scales before Hamiltonian Monte Carlo. The validation grid gives log-density errors below $10^{-4}$, and conditional scale moments can be averaged over the posterior draws.

The simulations show that the posterior recovers the parameters of the proposed model, with empirical coverage near the nominal level in most designs for the regression coefficients, and that the scale mixture substantially reduces the upper-quantile error relative to PALD in the Student-$t_3$ designs. In the tail-risk application on the Nifty Bank index, the Brazilian Ibovespa and the Indian rupee, series whose upper tails and central shapes depart from the Laplace benchmarks examined here, the proposed models forecast the conditional quantile more accurately than the ALD in 11 of 12 cases, with PSMALD improving on the ALD at the 99\% level in every series and provisional comparisons for two 99\% PALD fits; PSMALD gives the best density forecasts in every case, identifies the crisis days of 2008, 2013 and 2020 without supervision, and estimates a smaller fitted response of VaR to volatility.

Several extensions follow naturally. Estimating the tail index $a_\lambda$, possibly separately for each tail, would let the data choose the tail and might reduce the conservatism of PSMALD in calm periods. Letting $\lambda_t$ depend on lagged information would give a dynamic version of the model in the spirit of CAViaR \citep{engle2004} and time-varying Bayesian VaR models \citep{gerlach2011}. Finally, because the effective level $\tau_{\rm eff}$ can be estimated when a suitable reference error distribution is available, it could be used to recalibrate the quantile forecasts while keeping the density fit of PSMALD, combining the strengths of the framework's members in a single forecast.

\section*{Acknowledgments}
The authors acknowledge the use of ChatGPT and Claude for language refinement and assistance with presentation. AI-assisted tools also supported LaTeX formatting and technical consistency checks during manuscript preparation. The authors are responsible for verifying the mathematical arguments, computational results and final scientific interpretation.

\subsection*{Data and Code availability}
The methodology is implemented in the accompanying \texttt{R} package, which contains the model-fitting code, simulation and application workflows, and the extracted Mumbai series. The package is intended for public release on GitHub during the publication process. The data analysed in this paper are daily closing prices of the Nifty Bank index (\href{https://finance.yahoo.com/quote/\%5ENSEBANK/}{\texttt{\^{}NSEBANK}}), the Brazilian Ibovespa index (\href{https://finance.yahoo.com/quote/\%5EBVSP/}{\texttt{\^{}BVSP}}) and the Indian rupee--US dollar exchange rate (\href{https://finance.yahoo.com/quote/INR\%3DX/}{\texttt{INR=X}}), which are publicly available from Yahoo Finance (\url{https://finance.yahoo.com}). The series were downloaded on 23 November 2026, and the snapshot used in the paper is included in the accompanying R package \texttt{PowerSMAL}, so that all results can be reproduced without access to Yahoo Finance. The package contains the PALD, SMALD and PSMALD distribution functions, the Stan program, the backtests and the frozen data. The folder \texttt{analysis/} contains one script per table and figure (\texttt{tables/Table\_X.R}, \texttt{figures/Figure\_X.R}), four computation scripts whose saved results these scripts read, and a \texttt{README.md} that lists the run order. The package and the analysis code are available at \url{https://github.com/Niku2325}.

\subsection*{Supplementary materials availability}
Supplementary materials detailing all the proof and some further simulation and application results are available at the link:  \href{https://drive.google.com/file/d/1tWDm34SJEpWOrK4CJyeujtWduXBoDwo5/view?usp=sharing}{Supplementary\-Materials}.

\bibliographystyle{apalike}
\bibliography{references}
\end{document}